\documentclass[11pt, a4paper]{article}

\usepackage[utf8]{inputenc}
\usepackage[T1]{fontenc}
\usepackage{lmodern}
\usepackage[margin=2.2cm]{geometry}
\usepackage{amsmath, amssymb}
\usepackage{graphicx}
\usepackage{booktabs}
\usepackage{tabularx}
\usepackage{longtable}
\usepackage{multirow}
\usepackage{array}
\usepackage{setspace}
\usepackage{natbib}          
\usepackage{microtype}
\usepackage{float}
\usepackage{caption}
\usepackage{subcaption}
\usepackage{lineno}          
\usepackage{xcolor}
\usepackage{enumitem}

\usepackage{hyperref}
\definecolor{linkblue}{HTML}{0B5394}  
\hypersetup{
  colorlinks = true,
  citecolor  = linkblue,   
  linkcolor  = linkblue,   
  urlcolor   = linkblue,   
  filecolor  = linkblue
}

\graphicspath{{figures/}}
\makeatletter
\def\maxwidth{\ifdim\Gin@nat@width>\linewidth\linewidth\else\Gin@nat@width\fi}
\def\maxheight{\ifdim\Gin@nat@height>0.9\textheight0.9\textheight\else\Gin@nat@height\fi}
\makeatother
\setkeys{Gin}{width=\maxwidth,height=\maxheight,keepaspectratio}

\newcommand{\MerredinSdWp}{1.33}
\newcommand{\MerredinSdWpCI}{0.29 to 2.65}
\newcommand{\MerredinSdSpatial}{1.09}
\newcommand{\MerredinSdSpatialCI}{0.25 to 1.79}
\newcommand{\MerredinSdNugget}{0.81}
\newcommand{\MerredinSdNuggetCI}{0.04 to 1.46}
\newcommand{\MerredinArRow}{-0.11}
\newcommand{\MerredinArRowCI}{-0.76 to 0.51}
\newcommand{\MerredinArCol}{0.58}
\newcommand{\MerredinArColCI}{-0.05 to 0.96}
\newcommand{\MerredinDepthEightyVsThirty}{1.42}
\newcommand{\MerredinDepthEightyVsThirtyCI}{-0.51 to 3.13}
\newcommand{\MerredinDepthEightyVsThirtyP}{0.928}
\newcommand{\MerredinDepthOneTwentyVsThirty}{1.46}
\newcommand{\MerredinDepthOneTwentyVsThirtyCI}{-0.47 to 3.13}
\newcommand{\MerredinDepthOneTwentyVsThirtyP}{0.936}
\newcommand{\MerredinDepthOneTwentyVsEighty}{0.04}
\newcommand{\MerredinDepthOneTwentyVsEightyCI}{-1.75 to 1.79}

\newcommand{\MerredinDepthNSOneTwentyVsThirty}{1.65}
\newcommand{\MerredinDepthNSOneTwentyVsThirtyCI}{-0.13 to 2.97}

\newcommand{\MerredinTillThirty}{0.04}
\newcommand{\MerredinTillThirtyCI}{0.00 to 0.16}
\newcommand{\MerredinTillEighty}{0.10}
\newcommand{\MerredinTillEightyCI}{0.02 to 0.25}
\newcommand{\MerredinTillOneTwenty}{0.10}
\newcommand{\MerredinTillOneTwentyCI}{0.02 to 0.26}

\newcommand{\MerredinTillPOneTwentyVsThirty}{0.941}

\newcommand{\MerredinDepthWidthRThree}{2.07}
\newcommand{\MerredinDepthEstRThree}{1.46}
\newcommand{\MerredinDepthEstRThreeCI}{-0.47 to 3.13}
\newcommand{\MerredinDepthWidthRFour}{1.15}
\newcommand{\MerredinDepthEstRFour}{2.87}
\newcommand{\MerredinDepthEstRFourCI}{1.80 to 3.72}
\newcommand{\MerredinDepthWidthRatioRFour}{0.55}

\newcommand{\MerredinNCat}{7}
\newcommand{\MerredinNPlant}{899}

\newcommand{\WickepinSdWp}{0.33}
\newcommand{\WickepinSdWpCI}{0.02 to 0.88}
\newcommand{\WickepinSdSpatial}{0.41}
\newcommand{\WickepinSdSpatialCI}{0.02 to 0.89}
\newcommand{\WickepinSdNugget}{0.58}
\newcommand{\WickepinSdNuggetCI}{0.07 to 0.91}
\newcommand{\WickepinArRow}{-0.04}
\newcommand{\WickepinArRowCI}{-0.74 to 0.72}
\newcommand{\WickepinArCol}{0.28}
\newcommand{\WickepinArColCI}{-0.65 to 0.92}
\newcommand{\WickepinDepthEightyVsThirty}{0.80}
\newcommand{\WickepinDepthEightyVsThirtyCI}{-0.04 to 1.48}
\newcommand{\WickepinDepthEightyVsThirtyP}{0.970}
\newcommand{\WickepinDepthOneTwentyVsThirty}{1.75}
\newcommand{\WickepinDepthOneTwentyVsThirtyCI}{0.83 to 2.43}
\newcommand{\WickepinDepthOneTwentyVsThirtyP}{0.999}
\newcommand{\WickepinDepthOneTwentyVsEighty}{0.95}
\newcommand{\WickepinDepthOneTwentyVsEightyCI}{0.16 to 1.69}

\newcommand{\WickepinDepthNSOneTwentyVsThirty}{1.61}
\newcommand{\WickepinDepthNSOneTwentyVsThirtyCI}{0.76 to 2.24}

\newcommand{\WickepinTillThirty}{0.04}
\newcommand{\WickepinTillThirtyCI}{0.01 to 0.12}
\newcommand{\WickepinTillEighty}{0.11}
\newcommand{\WickepinTillEightyCI}{0.05 to 0.22}
\newcommand{\WickepinTillOneTwenty}{0.24}
\newcommand{\WickepinTillOneTwentyCI}{0.13 to 0.37}

\newcommand{\WickepinTillPOneTwentyVsThirty}{0.999}

\newcommand{\WickepinDepthWidthRThree}{0.91}
\newcommand{\WickepinDepthEstRThree}{1.75}
\newcommand{\WickepinDepthEstRThreeCI}{0.83 to 2.43}
\newcommand{\WickepinDepthWidthRFour}{0.57}
\newcommand{\WickepinDepthEstRFour}{1.89}
\newcommand{\WickepinDepthEstRFourCI}{1.40 to 2.38}
\newcommand{\WickepinDepthWidthRatioRFour}{0.63}

\newcommand{\WickepinNCat}{8}
\newcommand{\WickepinNPlant}{900}

\newcommand{\MerredinCVRZeroDiff}{-75.5}
\newcommand{\MerredinCVRZeroSe}{37.3}

\newcommand{\WickepinCVRZeroDiff}{-37.2}
\newcommand{\WickepinCVRZeroSe}{9.9}

\newcommand{\MerredinFamSequentialDiff}{-2.9}
\newcommand{\MerredinFamSequentialSe}{3.5}

\newcommand{\WickepinFamSequentialDiff}{-37.2}
\newcommand{\WickepinFamSequentialSe}{8.1}

\newcommand{\MerredinPseudoWidthDepthRZero}{1.77}
\newcommand{\MerredinPseudoWidthDepthRThree}{2.07}
\newcommand{\MerredinPseudoRatioDepth}{1.2}
\newcommand{\MerredinPseudoWidthVarietyRZero}{0.69}
\newcommand{\MerredinPseudoWidthVarietyRThree}{1.58}
\newcommand{\MerredinPseudoRatioVariety}{2.3}
\newcommand{\WickepinPseudoWidthDepthRZero}{0.83}
\newcommand{\WickepinPseudoWidthDepthRThree}{0.91}
\newcommand{\WickepinPseudoRatioDepth}{1.1}
\newcommand{\WickepinPseudoWidthVarietyRZero}{0.60}
\newcommand{\WickepinPseudoWidthVarietyRThree}{1.07}
\newcommand{\WickepinPseudoRatioVariety}{1.8}
\newcommand{\MerredinCVNuggetGain}{75.2}
\newcommand{\MerredinCVNuggetGainSe}{33.8}

\newcommand{\MerredinCVCorrGain}{-8.3}
\newcommand{\MerredinCVCorrGainSe}{7.6}
\newcommand{\MerredinCVThreeVFour}{-8.6}
\newcommand{\MerredinCVThreeVFourSe}{10.5}
\newcommand{\WickepinCVNuggetGain}{35.6}
\newcommand{\WickepinCVNuggetGainSe}{9.1}

\newcommand{\WickepinCVCorrGain}{0.9}
\newcommand{\WickepinCVCorrGainSe}{1.0}
\newcommand{\WickepinCVThreeVFour}{-0.7}
\newcommand{\WickepinCVThreeVFourSe}{1.8}
\newcommand{\MerredinCVPlotModelsMaxRatio}{1.1}

\newcommand{\WickepinCVPlotModelsMaxRatio}{0.8}

\newcommand{\ValCholErr}{3e-16}
\newcommand{\ValPmfErr}{3e-16}
\newcommand{\ValCovErr}{0.0111}
\newcommand{\ValCovErrTransposed}{1.16}
\newcommand{\ValBrmsMaxDiff}{0.056}
\newcommand{\ValCoverage}{0.99}
\newcommand{\ValNSim}{20}
\newcommand{\ValRhoRowBias}{-0.20}
\newcommand{\ValRhoColBias}{-0.17}
\newcommand{\ValRhoRowWidth}{1.04}
\newcommand{\ValSdSpatialBias}{-0.03}
\newcommand{\ValSdRepBias}{0.24}
\newcommand{\ValSdNuggetBias}{-0.02}
\newcommand{\ValBetaBias}{0.36}
\newcommand{\MerredinSensPrimary}{1.46}

\newcommand{\MerredinSensWideBeta}{3.54}

\newcommand{\MerredinSensFlatRho}{1.42}

\newcommand{\MerredinSensFlatRhoCol}{0.78}
\newcommand{\WickepinSensPrimary}{1.75}

\newcommand{\WickepinSensWideBeta}{2.04}

\newcommand{\WickepinSensFlatRho}{1.76}

\newcommand{\WickepinSensFlatRhoCol}{0.48}

\newcommand{\CrossSiteEightyVsThirtySameSign}{0.902}

\newcommand{\CrossSiteOneTwentyVsThirtySameSign}{0.936}

\newcommand{\CrossSiteOneTwentyVsEightySameSign}{0.519}
\newcommand{\ParetoKWorst}{78}
\newcommand{\ParetoKBlock}{16}
\newcommand{\NFold}{five}
\newcommand{\MaxRhat}{1.0096}
\newcommand{\MinEss}{727}
\newcommand{\MinEssTail}{1040}
\newcommand{\PrimaryMaxRhat}{1.0054}
\newcommand{\PrimaryMinEss}{1036}

\newcommand{\MerredinVhMaxAbs}{0.87}
\newcommand{\MerredinVhNExclZero}{1}
\newcommand{\WickepinVhMaxAbs}{0.86}
\newcommand{\WickepinVhNExclZero}{1}
\newcommand{\TotalDivergent}{19}

\newcommand{\MerredinScaleObsSratioDiff}{-2.9}
\newcommand{\MerredinScaleObsSratioSe}{3.5}

\newcommand{\MerredinScalePrinK}{5}

\newcommand{\MerredinScalePrinCumulDiff}{-0.1}
\newcommand{\MerredinScalePrinCumulSe}{1.9}

\newcommand{\MerredinLinkDiff}{-3.5}
\newcommand{\MerredinLinkSe}{6.5}
\newcommand{\MerredinHROneTwentyVsThirty}{0.23}

\newcommand{\WickepinScaleObsSratioDiff}{-37.2}
\newcommand{\WickepinScaleObsSratioSe}{8.1}

\newcommand{\WickepinScalePrinSratioDiff}{-36.0}
\newcommand{\WickepinScalePrinSratioSe}{7.9}

\newcommand{\WickepinLinkDiff}{-25.1}
\newcommand{\WickepinLinkSe}{5.3}
\newcommand{\WickepinHROneTwentyVsThirty}{0.32}

\title{Bayesian Ordinal Analysis of Cereal Growth Stage Data from Oat Trials in Western Australia}

\author{
  Zhanglong Cao\textsuperscript{1,*}
  \and Rose Megirian\textsuperscript{2}
  \and Matthew Nguyen\textsuperscript{2}
  \and Adam H. Sparks\textsuperscript{2}
}
\date{
  \small
  \textsuperscript{1}School of Electrical Engineering, Computing and Mathematical
Sciences, Curtin University, Perth, Western Australia, Australia\\[2pt]
  \textsuperscript{2}Centre for Crop and Disease Management,
    Curtin University, Perth, Western Australia, Australia\\[2pt]
  \textit{Corresponding author:}
    Zhanglong Cao (\href{mailto:zhanglong.cao@curtin.edu.au}{zhanglong.cao@curtin.edu.au})
}

\begin{document}

\maketitle

\begin{abstract}
Growth stage in cereal field trials is usually scored on the Zadoks decimal scale. The score is ordinal, yet it is often analysed as if it were continuous, which can lead to misleading conclusions. Ordinal analysis of growth stage data has received little attention from biometricians or agronomists, and it is not straightforward when several plants are scored in each plot of a split-plot design. We analysed Zadoks scores recorded four weeks after sowing on ten plants per plot in two deep-sowing oat trials in Western Australia, using a Bayesian hierarchical ordinal model that carries the full split-plot structure and an AR1\,$\times$\,AR1 spatial covariance. The implementation was validated on simulated data, and the candidate models were compared by five-fold cross-validation over plots. Adding a random effect at the experimental-unit (plot) level changed the uncertainty of the sub-plot contrasts substantially, while the main-plot contrasts changed little. How that plot-level heterogeneity was modelled mattered much less. At either site, cross-validation could not separate the candidate models that carried a plot-level term, whether or not they included an AR1\,$\times$\,AR1 field or the blocking terms, and the autoregressive parameters were only weakly identified on this layout. Deeper sowing (80 and 120~mm against 30~mm) went with more advanced development at four weeks in both trials. Sowing at 120~mm rather than 30~mm raised the predicted probability of tillering from \WickepinTillThirty{} to \WickepinTillOneTwenty{} at Wickepin and from \MerredinTillThirty{} to \MerredinTillOneTwenty{} at Merredin. The depth effects were stable whether or not the spatial field was fitted, and under both the cumulative and the sequential ordinal family.

\end{abstract}

\section{Introduction}
\label{sec:intro}

Herbicide timing, disease scouting and nitrogen scheduling all depend on how far a crop has developed, so growth stage is monitored routinely in cereal agronomy \citep{Zadoks1974}.  The Zadoks decimal scale describes that development in ten principal stages, each of them subdivided. The scale is ordinal and sequential, because a plant must pass through each stage before it reaches the next, but it is not chronological. A plant at three leaves (Z13) usually begins tillering (Z21) before its fourth leaf (Z14) is complete. The codes are labels for an ordered sequence, not measurements on an interval scale.

Despite this, growth stage data are commonly analysed by ANOVA or linear mixed models applied to the raw codes \citep{Stroup2015, Hartung2006}. Such analyses throw away the ordering constraint and can predict outside the valid range, and they cannot report the probability of being at or beyond a particular stage on a particular date \citep{Agresti2010, Harrell2015}. Ordinal regression is the natural alternative \citep{Peterson1989, burkner2019ordinal}, and it has been used for disease severity \citep{Shah2004} and for genomic prediction of categorical traits \citep{MontesinosLopez2015}. It keeps the ordering, respects the bounds, and returns the probability of each stage directly, so a treatment effect can be reported as the chance of having reached tillering rather than as a difference in average code. That is the form an agronomic recommendation takes.

The trials analysed here come from a deep sowing programme, and that setting determines what the growth stage data are being asked to show. In low-rainfall Mediterranean environments the sowing window is set by an opening rainfall event, and waiting for it costs yield through a shortened season \citep{Kirkegaard2014, Fletcher2015}. Sowing into stored subsoil moisture removes that dependence, but it asks the seedling to extend a coleoptile through 80 or 120~mm of soil rather than 30, which can delay emergence and thin the stand \citep{Rebetzke2007}. Whether the crop then makes up the ground, and how fast, is a question about early development rather than about final yield, and growth stage a few weeks after sowing is what answers it. A single census gives an interval-censored snapshot of that progress. It cannot separate emergence from subsequent development, but it does record where the population had reached, and it records it on the ordinal scale on which the trait is scored.

We see two further gaps in how such data are usually analysed, and the first concerns the unit of analysis. Growth stage is almost always assessed by scoring several plants per plot, and those plants are observational units while the plot is the experimental unit \citep{Hurlbert1984, Piepho2003, Casler2015}.  For a continuous response the standard remedy is to average within plot \citep{Millar2004}. That route is closed on an ordinal scale, because the mean of two Zadoks codes is not a Zadoks code \citep{Stroup2015}.  The alternative is to analyse the plant scores directly, which is correct only if a random effect at the plot level goes with it \citep{Bolker2009, Schielzeth2013}. How much difference that term makes in an ordinal analysis does not appear to have been quantified.

The second gap is spatial covariance. For continuous responses, separable autoregressive models on the row $\times$ column plot grid have been standard in variety testing and agronomy for three decades \citep{Cullis1991, Gilmour1997, Stefanova2009, Piepho2010spatial}. They are usually AR1\,$\times$\,AR1 with a nugget, and the same linear mixed model machinery now carries the analysis of large on-farm strip trials \citep{Stefanova2023}. Bayesian formulations go further and let the treatment effect itself vary across the field, estimating the spatial correlation and the treatment parameters jointly rather than in sequence \citep{cao2022bayesian}. The case for modelling space does not depend on the type of response. Neighbouring plots share soil and management, so their residuals are correlated, and treatments laid out in contiguous strips, as main plots always are, can be confounded with that correlation.

Ordinal rating scales are themselves common in agronomy and plant breeding, and the statistical questions they raise have been worked over before. \citet{Hartung2006} set out the case for treating such scales as ordinal rather than converting them to percentages, and \citet{Simko2011} address how ratings taken on different scales and in different experiments can be combined. Threshold models of this kind are now standard for genomic prediction of ordered categorical traits \citep{MontesinosLopez2015}, and ordinal regression has been applied to disease severity in field experiments \citep{Shah2004}. What these ordinal analyses have in common is that the plot grid plays no part in them. The random-effect structure follows the design, and whatever field trend remains is left in the error term. We are not aware of a field-trial ordinal analysis that carries a separable two-dimensional autoregressive field of the kind that is routine for continuous responses. 

Bayesian inference suits both gaps.  It returns full posteriors, so a statement such as ``the probability that this treatment is ahead'' follows directly. Hierarchical models also represent the nested strata of a split-plot design and carry uncertainty across them \citep{Gelman2013, cao2022bayesian}. We know of no standard package that fits a split-plot ordinal model with a two-dimensional autoregressive field, so the model has to be written and checked from scratch, and the checking matters as much as the fitting. We therefore follow a Bayesian workflow throughout, specifying priors, checking predictions, validating the implementation on simulated data and comparing models out of sample \citep{Gelman2026workflow, Vehtari2017, Cook2006}.

In this paper we specify, validate and apply a Bayesian hierarchical ordinal model with the full split-plot structure and a separable AR1\,$\times$\,AR1 spatial covariance. The data are Zadoks growth stage scores from two oat trials conducted in Western Australia in 2024 as part of a deep-sowing programme. The two sites are separate locations under independent management, which matters for the third objective below. A depth recommendation is only useful if it holds beyond the paddock it was measured in, and two separately managed sites are the smallest test of that. Our objectives are:
\begin{enumerate}[label=(\roman*)]
\item to specify, implement and validate a spatial Bayesian ordinal model for growth stage data recorded on a regular plot grid, with the experimental unit correctly represented;
\item to establish which random-effect, spatial and ordinal-family structures are supported by out-of-sample prediction, using a cross-validation scheme appropriate to subsampled data;
\item to estimate the effect of sowing depth on early development as decision-relevant posterior probabilities, and to assess whether that effect is consistent across two independently managed sites.
\end{enumerate}

\medskip
\noindent All three are addressed within a single Bayesian split-plot framework, with implementation checks and out-of-sample prediction as the main tools.

\section{Materials and methods}
\label{sec:methods}

\subsection{Experimental design and field layout}
\label{sec:design}

\begin{figure}[!htb]
\centering
\includegraphics[width=0.9\textwidth]{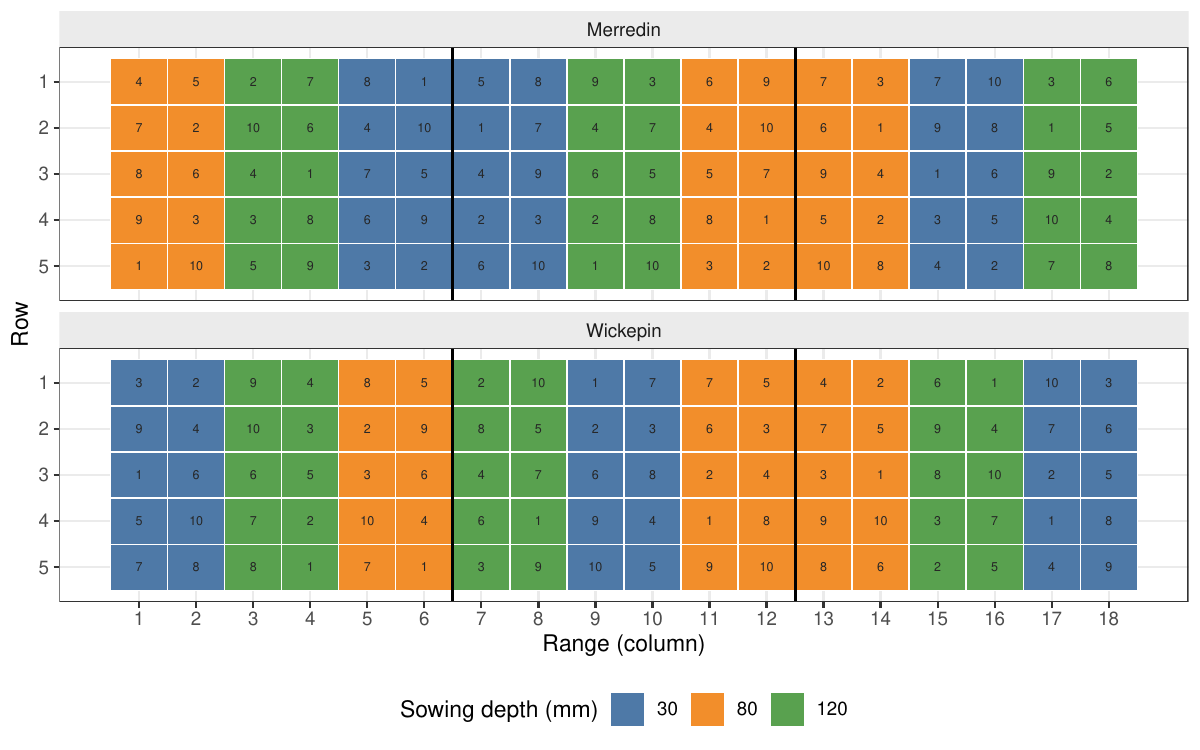}
\caption{Field layout of the two trials. Colour gives the main-plot sowing depth and the label is the variety--herbicide combination. Heavy vertical lines separate the three replicates, which are contiguous blocks of six ranges. Sowing depth is applied to contiguous two-range strips, so a smooth trend along the ranges could easily look like a depth effect, which is why an explicit spatial analysis is worth doing.}
\label{fig:layout}
\end{figure}

Two field trials were established in 2024 as part of a deep sowing project managed by the Facey Group, at Merredin and Wickepin (Western Australia). Both trials used split-plot designs on a rectangular grid of 90 plots arranged in 18 ranges and 5 rows (Figure~\ref{fig:layout}). The three replicates were contiguous blocks of six ranges (ranges 1--6, 7--12 and 13--18). Within each replicate, sowing depth was the main-plot, with levels 30, 80 and 120~mm. The ten variety--herbicide combinations (labelled V01--V10) were randomly allocated within each main plot. Every one of the 30 treatment combinations therefore occurred three times, once per replicate, and each occupied a single plot.

Two features of this layout drive the analysis. First, the main-plot stratum is real and has to be modelled, because sowing depth is replicated nine times, not ninety, and inference on depth that ignores the main-plot error term will be badly over-confident \citep{Milliken2009}. Second, sowing depth has to be applied in contiguous strips for practical reasons, and the replicate blocks are contiguous too, so any smooth spatial trend in the field would be structurally confounded with them, whether or not such a trend happens to be present. A spatial model is meant to separate a depth difference from the position of its strips in the field, but a contiguous-strip layout limits how well any model can make that separation. How far a strip layout can be relied on for that separation, and whether the strips are better placed systematically or at random, is a design question in its own right \citep{Cao2024}. We return to the tension in Section~\ref{sec:disc-spatial}.

Ten plants per plot were scored for growth stage four weeks after sowing following the Zadoks decimal scale \citep{Zadoks1974}, giving \WickepinNPlant{} plant-level observations at Wickepin and \MerredinNPlant{} at Merredin (one plant at Merredin was not scored and was omitted).

\subsection{The Zadoks scale and the observed category set}
\label{sec:zadoks}

The Zadoks decimal code describes ten principal developmental stages (0~=~germination, 1~=~seedling growth, 2~=~tillering, \ldots, 9~=~ripening), each subdivided into secondary stages \citep{Zadoks1974}. The stages relevant here are listed in Appendix~\ref{app:zadoks}. Development is not monotone in calendar time. On completing three leaves (Z13) a plant begins tillering (Z21) before completing its fourth leaf. The observed sequence therefore contains a jump from the leaf-emergence phase to the tillering phase.

An ordinal model estimates one threshold per adjacent pair of categories, and a category that is never observed carries no information about its own threshold. In the sequential family every plant that reaches such a category passes straight through it, so its conditional probability of advancing is one and the threshold diverges; in the cumulative family the two thresholds that bound it are driven together. Either way the parameter is fixed by the prior rather than by the data, while the fitted probabilities of the categories that were observed are unchanged. We therefore defined the categories by what was observed at each site. At Merredin the ordered set has \MerredinNCat{} levels (Z00, Z01, Z11, Z12, Z13, Z21, Z22) and at Wickepin \WickepinNCat{} levels (Z11--Z15, Z21, Z23, Z24). Counts are given in Table~\ref{tab:descriptive}. We omitted the intermediate codes that did not occur: Z03, Z05, Z07, Z09 and Z10 at Merredin, and Z22 at Wickepin. A plant does of course pass through those stages, and omitting them does not deny it. An ordinal model uses only the ordering of the categories, so the developmental sequence is preserved either way, and what is dropped is a set of parameters the data cannot inform.

\subsection{Ordinal model families}
\label{sec:families}

Let $Y_i \in \{1, \ldots, K\}$ denote the growth stage of plant $i$, ordered from least to most advanced, let $\eta_i$ be a linear predictor (Section~\ref{sec:linpred}) common to every $\alpha_k$, and let $\alpha_k$, $k = 1, \ldots, K-1$, denote a category-specific intercept, interpreted differently by family below. We considered three families, the three in standard use for ordinal responses \citep{Agresti2010, burkner2019ordinal}, each written so that a larger $\eta$ always means more advanced development:

\begin{enumerate}
\item \textbf{Cumulative (proportional odds):} $P(Y_i > k) = \mathrm{logit}^{-1}(\alpha_k + \eta_i)$ with $\alpha_1 > \cdots > \alpha_{K-1}$. A single latent variable is cut at $K-1$ ordered thresholds; treatment effects shift all thresholds equally.
\item \textbf{Sequential (continuation ratio):}
    \begin{equation}
      P(Y_i > k \mid Y_i \geq k) = \mathrm{logit}^{-1}(\alpha_k + \eta_i),
      \label{eq:sratio}
    \end{equation}
so that $P(Y_i = k) = \bigl[\prod_{j<k} \mathrm{logit}^{-1}(\alpha_j + \eta_i)\bigr] \bigl[1 - \mathrm{logit}^{-1}(\alpha_k + \eta_i)\bigr]$ and $P(Y_i = K) = \prod_{j<K} \mathrm{logit}^{-1}(\alpha_j + \eta_i)$. Here the $\alpha_k$ are unconstrained, and each is a transition-specific parameter describing conditional advancement at the census.
\item \textbf{Adjacent-category:} $P(Y_i = k) \propto \exp\!\bigl(\sum_{j<k} (\alpha_j + \eta_i)\bigr)$, normalised over $k$; local stage-to-stage ratios, intermediate in flexibility.
\end{enumerate}

The sequential model is the natural candidate for developmental data, because it conditions on the plant having reached stage $k$ and asks whether it advanced \citep{Tutz1991, burkner2019ordinal}. It also gives the Z13~$\to$~Z21 step a parameter of its own that reads directly as the conditional probability of making that step, so a transition of qualitatively different difficulty shows up in a single $\alpha_k$. The adjacent-category family likewise leaves $\alpha$ unordered, but its $\alpha_k$ describe the ratio of neighbouring category probabilities rather than conditional advancement. With a logit link the sequential model is a discrete-time hazard model, in which Equation~\eqref{eq:sratio} is the conditional probability of advancing at step $k$ having reached it and $\alpha_k$ is the log-odds of that hazard.  A growth stage scored at a fixed census date is interval-censored progress through such a sequence, so the hazard form matches the way the data are generated.  The cumulative model instead assumes a continuous latent developmental age, plausibly accumulated thermal time, coarsened into the recorded stages. That is a coherent alternative too, but it requires proportional odds, a common odds ratio at every cut point.

We therefore take the sequential family as the model of record at both sites, on the mechanistic grounds just given, and this is not something we ask cross-validation to overturn. What cross-validation can settle is a different question. In all three families the treatment-effect coefficients $\boldsymbol{\beta}$ (Section~\ref{sec:linpred}) are common to all thresholds, which is the proportional form of each, and that is an assumption rather than a consequence of the mechanism. A family that loses on predictive accuracy is telling us that its own constant-effect assumption is strained, not that development proceeds by some other route. We read the comparison in Section~\ref{sec:modelcomp} in that light, and report the cumulative fits alongside as a sensitivity check.

\subsection{Hierarchical and spatial linear predictor}
\label{sec:linpred}
\label{sec:units}

Ten plants were scored per plot, and how they enter the model is a key specification decision. The experimental unit is the plot, that being the smallest unit to which a treatment was independently randomised, and the observational unit is the plant. Scoring ten plants per plot is subsampling, which pins down the plot's developmental state more precisely but adds no replication of the treatment \citep{Hurlbert1984, Piepho2003, Casler2015}. An analysis entering the 900 plant scores without a plot-level random effect implicitly asserts that they are 900 independent responses to treatment, understating standard errors as though the trial had 900 experimental units rather than 90.

Averaging to a plot mean is the usual remedy for continuous responses \citep{Millar2004}, but it is not available on an ordinal scale, because Zadoks codes label an ordered sequence and the mean of Z13 and Z21 has no developmental meaning \citep{Agresti2010, Stroup2015}. What does work is to retain the plant-level data and add a random effect at the level of the experimental unit \citep{Bolker2009, Schielzeth2013}, the term $e_p$ in Equation~\eqref{eq:eta}. All ten plants in plot $p$ share the same value $\eta_p$ of the linear predictor (Equation~\eqref{eq:eta} below), so that term induces the within-plot correlation created by subsampling, while the multinomial form of the ordinal likelihood accounts for plant-to-plant variation around it. A separate plant-within-plot effect would have one observation per level and is neither identified nor needed.

Two points follow from the split-plot structure. First, each factor must be evaluated against its own error stratum \citep{Piepho2003}. Sowing depth is a main-plot factor, with its own random effect $w$ and standard deviation $\sigma_{\mathrm{mp}}$ defined below (Equations~\eqref{eq:eta}--\eqref{eq:re}), so a model that retains $w$ but drops $e_p$ still treats the depth contrast roughly correctly. Variety--herbicide is a sub-plot factor whose error stratum is the plot itself, and omitting $e_p$ leaves it without an appropriate error stratum (Section~\ref{sec:res-pseudo}). Second, the ten plants recorded per plot are exchangeable sample identifiers rather than a cross-plot grouping factor. Indexing a term by plant number would pool arbitrary plants across plots \citep{Schielzeth2013}.

We therefore define the linear predictor at the plot level. For plot $p$ at grid position (row $r(p)$, range $c(p)$), in replicate $\mathrm{rep}(p)$ and main plot $\mathrm{mp}(p)$:
\begin{equation}
  \eta_p = \mathbf{x}_p^\top \boldsymbol{\beta}
         + u_{\mathrm{rep}(p)}
         + w_{\mathrm{mp}(p)}
         + s_{r(p),\, c(p)}
         + e_p,
  \label{eq:eta}
\end{equation}
and $\eta_i = \eta_{p(i)}$ for every plant $i$ in plot $p$. Here $\mathbf{x}_p$ holds the treatment contrasts (Section~\ref{sec:fixed}), $u$ is the replicate effect, $w$ the main-plot (replicate $\times$ sowing depth) effect that forms the split-plot error stratum, $s$ the spatial field, and $e_p$ an independent plot-level nugget:
\begin{equation}
  u_j \sim \mathcal{N}(0, \sigma_{\mathrm{rep}}^2), \qquad
  w_j \sim \mathcal{N}(0, \sigma_{\mathrm{mp}}^2),  \qquad
  e_p \sim \mathcal{N}(0, \sigma_e^2).
  \label{eq:re}
\end{equation}
There is no intercept column in $\mathbf{x}_p$, because the thresholds $\alpha_k$ absorb it.

The spatial field $\mathbf{S}$, an $N_{\mathrm{row}} \times N_{\mathrm{col}}$ matrix with entries $s_{r,c}$, follows a separable AR1\,$\times$\,AR1 process constructed as
\begin{equation}
  \mathbf{S} = \sigma_s\, \mathbf{L}_r\, \mathbf{Z}\, \mathbf{L}_c^\top,
  \label{eq:spatial}
\end{equation}
where $\mathbf{Z}$ has independent standard normal entries and $\mathbf{L}_r$, $\mathbf{L}_c$ are the lower Cholesky factors of the AR1 correlation matrices $\boldsymbol{\Sigma}_r[i,j] = \rho_r^{|i-j|}$ and $\boldsymbol{\Sigma}_c[i,j] = \rho_c^{|i-j|}$. This construction gives
\begin{equation}
  \mathrm{Cov}(s_{i,j},\, s_{k,l})
    = \sigma_s^2\, \rho_r^{|i-k|}\, \rho_c^{|j-l|},
  \label{eq:cov}
\end{equation}
that is, writing $\mathrm{vec}(\cdot)$ for column-major vectorisation (a matrix's columns stacked into a single vector), $\mathrm{vec}(\mathbf{S}) \sim \mathcal{N}(\mathbf{0},\, \sigma_s^2\, \boldsymbol{\Sigma}_c \otimes \boldsymbol{\Sigma}_r)$ as rows nested within columns, with marginal standard deviation $\sigma_s$ at every plot. 

Two implementation details matter. First, the Cholesky factor of an AR1 correlation matrix is available in closed form,
\begin{equation}
  L_{11} = 1, \quad
  L_{i1} = \rho^{\,i-1}, \quad
  L_{ii} = \sqrt{1-\rho^2}\ (i \geq 2), \quad
  L_{ij} = \rho^{\,i-j}\sqrt{1-\rho^2}\ (i > j \geq 2),
  \label{eq:chol}
\end{equation}
so no numerical decomposition and no diagonal jitter are needed and $\rho$ may approach $\pm 1$ safely. Second, we allow $\rho_r$ and $\rho_c$ the full range $(-1, 1)$. Negative autocorrelation between adjacent plots is a real phenomenon in field trials, arising from inter-plot competition, and restricting the parameters to be positive would build in an assumption the data should be allowed to contradict.

The nugget $e_p$ and the spatial field are both plot-level terms, and they are distinguished only through $\rho_r$ and $\rho_c$, since as $\rho_r, \rho_c \to 0$ the field degenerates to independent plot effects. Including both is nevertheless standard \citep{Gilmour1997}, because without a nugget the autoregressive parameters are forced to absorb plot-level noise, and the model ladder below tests whether the data support the separation.

\subsection{Fixed-effect structure}
\label{sec:fixed}

We considered three treatment structures: sowing depth alone, depth plus the variety--herbicide combination additively, and the saturated depth $\times$ variety--herbicide interaction. The saturated structure is a reparameterisation of a single 30-level treatment factor, the specification used in earlier analyses of these trials. The two give the same fit, but only the saturated form has contrasts that can be read directly as depth and variety effects. We centred the design matrix before fitting, so that the threshold prior applies to an average treatment rather than to whichever level happens to be the reference. The treatment contrasts themselves are unaffected by the centring. We average all depth contrasts reported below over the ten variety--herbicide combinations.

\subsection{Prior distributions}
\label{sec:priors}

We used weakly informative priors throughout, following \citet{gelman2006prior} and \citet{McElreath2020}:
\begin{align}
  \alpha_k                            &\sim \text{Student-}t(3, 0, 2.5), \label{eq:prior1}\\
  \beta_j                             &\sim \mathcal{N}(0, 1),           \label{eq:prior2}\\
  \sigma_{\mathrm{rep}},\ \sigma_{\mathrm{mp}},\ \sigma_s,\ \sigma_e
                                      &\sim \text{Exponential}(1),       \label{eq:prior3}\\
  (\rho_r + 1)/2,\ (\rho_c + 1)/2     &\sim \text{Beta}(2, 2),           \label{eq:prior4}\\
  Z_{ij}                              &\sim \mathcal{N}(0, 1).           \label{eq:prior5}
\end{align}
The Student-$t$ prior on thresholds is heavy-tailed, so it tolerates the very unequal category spacings that the Zadoks scale produces. The $\mathcal{N}(0,1)$ prior on log-odds contrasts places about 95\% of its mass within $\pm 2$ logits, which is generous for a four-week developmental difference while still regularising the 29 contrasts of the saturated model. The Beta$(2,2)$ prior on the rescaled autocorrelations is symmetric about $\rho = 0$, has zero density at $\pm 1$, and does not presume positive spatial correlation. We gave the random effects non-centred parameterisations.

\subsection{Model comparison}
\label{sec:modelcomp}

We compared models by out-of-sample predictive accuracy in three stages: random-effect and spatial structure, ordinal family, and fixed-effect structure (Tables~\ref{tab:cv-random}--\ref{tab:cv-fixed}). We prespecified R3 and the sequential family for substantive inference, and treat the family and fixed-effect comparisons as sensitivity checks. The random-effect ladder was
\begin{itemize}[noitemsep]
\item[R0] replicate $+$ main plot, without a plot-level term (a plant-level ordinal comparison model);
\item[R1] R0 $+$ independent plot nugget;
\item[R2] R0 $+$ AR1\,$\times$\,AR1 field;
\item[R3] R0 $+$ AR1\,$\times$\,AR1 field $+$ nugget;
\item[R4] AR1\,$\times$\,AR1 field $+$ nugget, with the blocking terms removed, to evaluate whether spatial covariance can substitute for design-based blocking.
\end{itemize}

Because the ten plants scored in a plot are subsamples sharing soil, sowing depth, seed lot and micro-environment, cross-validation that holds out individual plants merely evaluates within-plot interpolation. We therefore held out entire plots as the cross-validation unit \citep{Merkle2019}.

Leave-one-plot-out cross-validation can in principle be approximated from a single fit by Pareto-smoothed importance sampling (PSIS-LOO) \citep{Vehtari2017, Vehtari2024psis}, but the approximation proved unstable in this setting. Removing a plot removes all direct information about that plot's own nugget and its own value of the spatial field, which makes the importance weights heavy-tailed. The Pareto $\hat{k}$ diagnostic therefore exceeded 0.7 for a substantial fraction of plots, most of all in the more flexible models (Table~\ref{tab:pareto-k}).

We therefore implemented exact \NFold-fold cross-validation over plots. Each model is refitted \NFold{} times with one fold of plots excluded from the likelihood, and the held-out log predictive density $\log p(\mathbf{y}_p \mid \mathbf{y}_{-\mathrm{fold}})$ of each plot is evaluated by averaging over the posterior draws of that fit. Summed over plots, this gives the expected log predictive density (elpd), so larger (less negative) values indicate better out-of-sample prediction. Folds were stratified by treatment cell (sowing depth crossed with variety--herbicide) across replicates, so that treatments were represented evenly across folds. Differences are reported with the standard error of the paired plot-wise differences, and a difference of more than two standard errors is taken as clear evidence. Posterior predictive checks on category frequencies accompany the numerical comparison.

R3 keeps the design-based blocking terms and adds the spatial field. We retained it for substantive inference because it was within one standard error of the best-ranked random-effect structure, and did not treat predictive equivalence as a reason to drop a randomisation stratum. We also retained the common interaction parameterisation for cross-site comparability, even though simpler fixed-effect structures ranked better in the exploratory CV.

\subsection{Computation}
\label{sec:computation}

We wrote the model of Sections~\ref{sec:families}--\ref{sec:priors} directly in Stan \citep{Carpenter2017} and fitted it through \texttt{rstan} in \texttt{R} \citep{Rcore2024}. A single Stan program serves the whole ladder. The ordinal family and each random-effect term are switched on or off from the data block, so every model in the comparison shares identical likelihood code and only the intended component changes. Plants within a plot are exchangeable and share $\eta_p$, so the likelihood is evaluated once per (plot, category) pair, weighted by the observed plant counts, rather than once per plant. That reduces the number of likelihood evaluations by an order of magnitude at no cost in fidelity.

Posterior inference used the No-U-Turn Sampler \citep{Hoffman2014}, an adaptive form of Hamiltonian Monte Carlo \citep{Neal2011}, running four chains of 8\,000 iterations (2\,000 warm-up, thinned by three) with \texttt{adapt\_delta}~$=0.99$. The cross-validation refits, which only need the held-out predictive density, used four chains of 2\,000 iterations (1\,000 warm-up). We assessed convergence for every model by the rank-normalised $\hat{R}$, the bulk and tail effective sample sizes \citep{Vehtari2021rhat} and the number of divergent transitions (Table~\ref{tab:diagnostics}). The \texttt{generated quantities} block returns pointwise log-likelihoods at both the plant and the plot level, and posterior predictive replicates.

Reported stage probabilities are for a plant in a new plot. At each posterior draw we integrate the category probabilities over the replicate, main-plot, spatial and nugget random effects, whose sum is normal with variance $\sigma_{\mathrm{rep}}^2 + \sigma_{\mathrm{mp}}^2 + \sigma_s^2 + \sigma_e^2$. The AR1\,$\times$\,AR1 field has marginal variance $\sigma_s^2$ at every plot. The integration uses 40-node equal-weight quantile quadrature at each draw rather than one random draw per posterior sample. The spread across draws therefore reflects parameter uncertainty alone, and the intervals are credible intervals for a well-defined quantity. We average depth contrasts over the ten variety--herbicide combinations on the latent scale, and stage probabilities on the probability scale.

\subsection{Validation of the implementation}
\label{sec:validation}

A spatial ordinal model written from scratch should not be trusted on the strength of its output alone \citep{Gelman2026workflow, Cook2006}. We ran four checks before extracting any scientific result, and machine-readable output from all four accompanies the analysis code.

\begin{enumerate}
\item \textbf{Unit tests of the Stan functions.} The closed-form AR1 Cholesky factor of Equation~\eqref{eq:chol} was compared with a numerical decomposition of $\boldsymbol{\Sigma}(\rho)$ for $\rho \in \{-0.9, \ldots, 0.999\}$, and the sequential category probabilities were compared with an independent \texttt{R} implementation over randomly drawn thresholds and linear predictors.
\item \textbf{Induced covariance.} The claim of Equation~\eqref{eq:cov} was verified by Monte Carlo. We generated 200\,000 realisations of $\sigma_s \mathbf{L}_r \mathbf{Z} \mathbf{L}_c^\top$ and compared their empirical covariance with $\sigma_s^2\,\boldsymbol{\Sigma}_c \otimes \boldsymbol{\Sigma}_r$. This check matters, because the transposed Kronecker ordering $\boldsymbol{\Sigma}_r \otimes \boldsymbol{\Sigma}_c$ is the intuitive one to write down and is wrong.
\item \textbf{Agreement with an independent engine.} The non-spatial version of each of the three ordinal families was fitted with both the Stan program and \texttt{brms} \citep{Burkner2017} and the posteriors compared.
\item \textbf{Parameter recovery.} Twenty data sets were simulated from the full model, with known $\boldsymbol{\beta}$, thresholds, variance components and autocorrelations, on the real 5\,$\times$\,18 field layout with ten plants per plot, and refitted. Bias and the coverage of 95\% credible intervals were recorded for every parameter.
\end{enumerate}

\section{Results}
\label{sec:results}

\subsection{Validation}
\label{sec:res-validation}

All four checks passed; we give the full output with the analysis code. The closed-form Cholesky factor of Equation~\eqref{eq:chol} reproduced a numerical decomposition to machine precision (largest absolute discrepancy \ValCholErr{}), and the sequential category probabilities matched an independent implementation to \ValPmfErr{} while summing to one. The Monte Carlo covariance of the Kronecker construction agreed with Equation~\eqref{eq:cov} to \ValCovErr{}; the transposed ordering $\boldsymbol{\Sigma}_r \otimes \boldsymbol{\Sigma}_c$, which is the intuitive one to write down, was out by \ValCovErrTransposed{}, two orders of magnitude more. Fitted against \texttt{brms}, the Stan program reproduced all three families to within \ValBrmsMaxDiff{} on the depth contrasts. The comparison also turned up a parameterisation detail worth recording. \texttt{brms} centres the population-level design matrix before applying the threshold prior, so a prior that is nominally on the thresholds in fact acts on the thresholds of an average treatment.  With an uncentred design matrix the two engines differed by about 0.2 logits for the cumulative family despite identical likelihoods; matching the centring removed the discrepancy, and the centred parameterisation was adopted (Section~\ref{sec:fixed}).

Parameter recovery on the real field layout (Figure~\ref{fig:recovery}) gave 95\% interval coverage of \ValCoverage{} across \ValNSim{} simulated data sets, with small bias in the thresholds and plot-level variance components (spatial standard deviation to within \ValSdSpatialBias{}, nugget to within \ValSdNuggetBias{}). We name three biases here, because they come from the design and the priors rather than from the implementation. Treatment contrasts shrank toward zero by up to \ValBetaBias{}, which is what an $\mathcal{N}(0,1)$ prior on 29 contrasts should do, so the reported effects are conservative; Section~\ref{sec:sensitivity} checks this. The replicate standard deviation was biased upward by \ValSdRepBias{}, because with three replicates $\sigma_{\mathrm{rep}}$ is largely prior-driven. The autocorrelations were also pulled toward zero (\ValRhoRowBias{} for $\rho_r$, \ValRhoColBias{} for $\rho_c$), with intervals averaging \ValRhoRowWidth{} wide. These data were simulated from known autocorrelations, so this says something about what a 5\,$\times$\,18 grid with contiguous replicates can identify, and it shapes how we read the spatial component below.

\subsection{Descriptive results}
\label{sec:descriptive}

\begin{figure}[htbp]
\centering
\includegraphics[width=\textwidth]{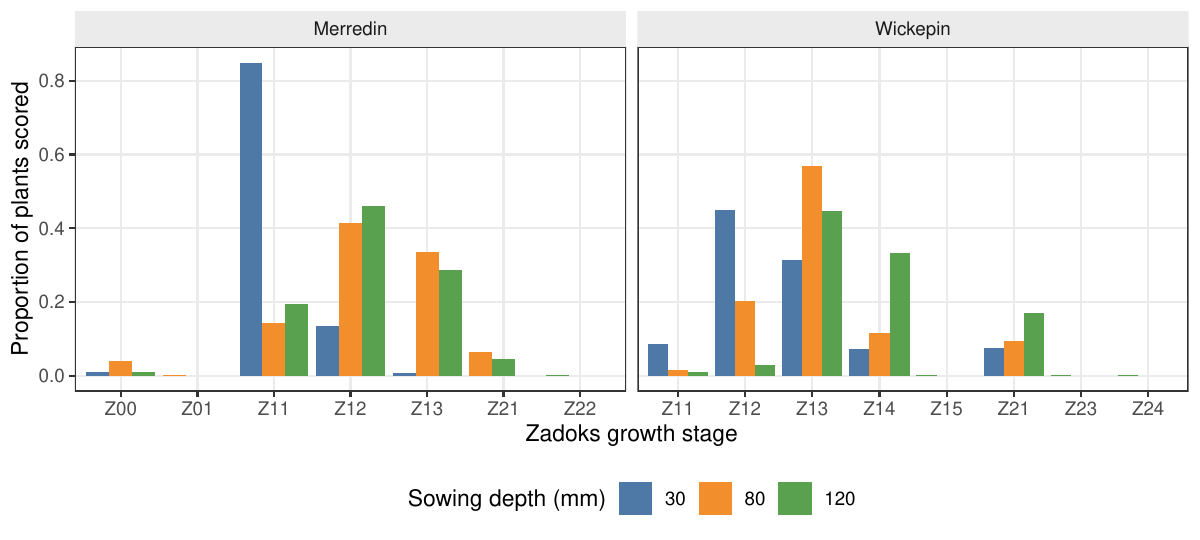}
\caption{Observed distribution of Zadoks growth stages by sowing depth, four weeks after sowing. Only stages observed at a site are shown.}
\label{fig:stagedist}
\end{figure}

Table~\ref{tab:descriptive} and Figure~\ref{fig:stagedist} summarise the observed distributions.

\begin{table}[htbp]
\centering
\caption{Number of plants scored (percentage within sowing depth in parentheses) at each observed Zadoks stage, by site and sowing depth. Stages never observed at a site are omitted; these categories are not estimable in an ordinal model and were excluded from the analysis.}
\label{tab:descriptive}
\small
\begin{tabular}{llrrr}
\toprule
Site & Zadoks stage & 30 mm & 80 mm & 120 mm \\
\midrule
Merredin & Z00 & 3 (1.0) & 12 (4.0) & 3 (1.0) \\
Merredin & Z01 & 0 (0.0) & 1 (0.3) & 0 (0.0) \\
Merredin & Z11 & 254 (84.9) & 43 (14.3) & 58 (19.3) \\
Merredin & Z12 & 40 (13.4) & 124 (41.3) & 138 (46.0) \\
Merredin & Z13 & 2 (0.7) & 101 (33.7) & 86 (28.7) \\
Merredin & Z21 & 0 (0.0) & 19 (6.3) & 14 (4.7) \\
Merredin & Z22 & 0 (0.0) & 0 (0.0) & 1 (0.3) \\
Wickepin & Z11 & 26 (8.7) & 5 (1.7) & 3 (1.0) \\
Wickepin & Z12 & 135 (45.0) & 61 (20.3) & 9 (3.0) \\
Wickepin & Z13 & 94 (31.3) & 171 (57.0) & 134 (44.7) \\
Wickepin & Z14 & 22 (7.3) & 35 (11.7) & 100 (33.3) \\
Wickepin & Z15 & 0 (0.0) & 0 (0.0) & 1 (0.3) \\
Wickepin & Z21 & 23 (7.7) & 28 (9.3) & 51 (17.0) \\
Wickepin & Z23 & 0 (0.0) & 0 (0.0) & 1 (0.3) \\
Wickepin & Z24 & 0 (0.0) & 0 (0.0) & 1 (0.3) \\
\bottomrule
\end{tabular}
\end{table}

At Merredin, development four weeks after sowing was concentrated in Z11--Z13, and the effect of sowing depth was pronounced. Of the plants sown at 30~mm, 84.9\% were still at Z11, against 14.3\% at 80~mm and 19.3\% at 120~mm. Plants sown at 80 and 120~mm were mostly at Z12 or Z13 (75.0\% and 74.7\% respectively), and only the deeper treatments produced any tillering (6.3\% at 80~mm and 5.0\% at 120~mm reached Z21 or beyond, against none at 30~mm). A small number of plants recorded Z00 or Z01, indicating delayed or failed germination; these occurred at all three depths but were most frequent at 80~mm (4.3\%).

At Wickepin the whole trial was further advanced and the same ordering held. The modal stage was Z12 at 30~mm (45.0\%), Z13 at 80~mm (57.0\%) and Z13 at 120~mm (44.7\%) with a further 33.3\% at Z14. The proportion of plants that had begun tillering rose monotonically with depth, from 7.7\% at 30~mm to 9.3\% at 80~mm and 17.7\% at 120~mm.

One feature of the raw data bears on the spatial analysis. The field means by range vary a good deal at both sites, but the pattern tracks the sowing depth strips rather than a smooth gradient, and the 30~mm strips are the least advanced wherever in the field they sit, in all three replicates and at both sites. The treatment ordering therefore repeats at three widely separated positions in the field, which we read as informal but useful evidence that the depth effect is not a spatial artefact.

\subsection{Model comparison}
\label{sec:modelcomp-results}

The Pareto $\hat{k}$ diagnostics (Table~\ref{tab:pareto-k}) confirmed that PSIS-LOO was unreliable for held-out plots in this setting. The block-structured model R0 exceeded $\hat{k} = 0.7$ for \ParetoKBlock{} of the 90 plots, and models incorporating plot-level nugget or spatial fields exceeded the threshold for up to \ParetoKWorst{} plots. Models were therefore compared by exact \NFold-fold cross-validation over plots.

Table~\ref{tab:cv-random} gives the Stage~1 comparison of random-effect and spatial structures. Two consistent patterns run through both sites.

\begin{table}[htbp]
\centering
\caption{Exact five-fold cross-validation over plots for the random-effect and spatial structure, holding the ordinal family (sequential) and the saturated depth $\times$ variety fixed effects constant. Models are ordered within site from best to worst; the difference is measured against the best model at that site. Higher (less negative) elpd indicates better out-of-sample prediction of a plot the model has not seen.}
\label{tab:cv-random}
\small
\begin{tabular}{llrrrr}
\toprule
Site & Model & elpd$_{5\mathrm{-fold}}$ & SE & Difference & SE of difference \\
\midrule
Merredin & R4: AR1$\times$AR1 + nugget, no blocking & -889.8 & 51.1 & 0.0 & 0.0 \\
Merredin & R1: R0 + plot nugget & -890.0 & 52.8 & -0.3 & 4.8 \\
Merredin & R2: R0 + AR1$\times$AR1 & -895.8 & 55.1 & -6.0 & 7.4 \\
Merredin & R3: R0 + AR1$\times$AR1 + nugget & -898.4 & 56.9 & -8.6 & 10.5 \\
Merredin & R0: rep + main-plot & -965.3 & 72.7 & -75.5 & 37.3 \\ \midrule
Wickepin & R4: AR1$\times$AR1 + nugget, no blocking & -1158.0 & 32.8 & 0.0 & 0.0 \\
Wickepin & R3: R0 + AR1$\times$AR1 + nugget & -1158.7 & 33.2 & -0.7 & 1.8 \\
Wickepin & R2: R0 + AR1$\times$AR1 & -1158.8 & 33.2 & -0.9 & 2.0 \\
Wickepin & R1: R0 + plot nugget & -1159.6 & 33.1 & -1.6 & 2.2 \\
Wickepin & R0: rep + main-plot & -1195.2 & 34.2 & -37.2 & 9.9 \\
\bottomrule
\end{tabular}
\end{table}

First, the classical split-plot model R0, with no plot-level random term, was the worst at both sites, with elpd differences from the best model of \MerredinCVRZeroDiff{} (SE \MerredinCVRZeroSe{}) at Merredin and \WickepinCVRZeroDiff{} (SE \WickepinCVRZeroSe{}) at Wickepin. Adding an independent plot nugget recovered essentially all of that, worth \MerredinCVNuggetGain{} (SE \MerredinCVNuggetGainSe{}) at Merredin and \WickepinCVNuggetGain{} (SE \WickepinCVNuggetGainSe{}) at Wickepin.

Second, making that plot-level term spatially correlated bought nothing further at either site. Against R1, the model with an independent nugget, the full AR1\,$\times$\,AR1 field of R3 was worth \MerredinCVCorrGain{} (SE \MerredinCVCorrGainSe{}) at Merredin and \WickepinCVCorrGain{} (SE \WickepinCVCorrGainSe{}) at Wickepin, which is within about one standard error of zero at both sites and negative at Merredin. What the data ask for is plot-level heterogeneity, not a particular correlation structure for it. This is consistent with the recovery study, which found the autocorrelations only weakly identified on a 5\,$\times$\,18 grid (Section~\ref{sec:res-validation}), and with the posterior intervals for $\rho_r$ and $\rho_c$ reported in Section~\ref{sec:varcomp}.

R4, which keeps the spatial field and the nugget but drops the replicate and main-plot terms, ranked first at both sites. The ranking should not be read as a result. None of the six comparisons among the four structures that carry a plot-level term reaches two standard errors at either site, the largest being \MerredinCVPlotModelsMaxRatio{} standard errors at Merredin and \WickepinCVPlotModelsMaxRatio{} at Wickepin. Measured as elpd of R3 minus R4, so that a negative value favours R4, the gap is \MerredinCVThreeVFour{} (SE \MerredinCVThreeVFourSe{}) at Merredin and \WickepinCVThreeVFour{} (SE \WickepinCVThreeVFourSe{}) at Wickepin.

What Stage~1 establishes is therefore narrower than the ordering in Table~\ref{tab:cv-random} suggests. A plot-level term is essential, and the evidence for that is strong at both sites. How it is modelled (independent or spatially correlated, with or without the design-based blocking terms) makes no difference that these data can detect to the prediction of a plot the model has not seen. Using a spatial field and nugget in place of design-based blocking is common practice in field-trial analysis \citep{Cullis1991, Gilmour1997}; that it was not measurably better here is a statement about what 90 plots on a 5-row grid can resolve, not a defect in the substitution itself.

Predictive equivalence is not a licence to drop the strata, and this data set gives a concrete example of why. Sowing depth is a main-plot factor, applied to nine 2-range $\times$ 5-row strips, so under R4 it has no error stratum of its own and is judged against plot-level noise instead. The variety-averaged 120 versus 30~mm contrast is \MerredinDepthEstRThree{} (95\% CI \MerredinDepthEstRThreeCI{}) under R3 at Merredin but \MerredinDepthEstRFour{} (\MerredinDepthEstRFourCI{}) under R4, and the mean width of the depth contrasts falls from \MerredinDepthWidthRThree{} to \MerredinDepthWidthRFour{}. At Wickepin the same comparison runs \WickepinDepthEstRThree{} (\WickepinDepthEstRThreeCI{}) against \WickepinDepthEstRFour{} (\WickepinDepthEstRFourCI{}), with widths \WickepinDepthWidthRThree{} and \WickepinDepthWidthRFour{}. At Merredin that is the difference between an interval that covers zero and one that does not, obtained by removing the stratum against which the factor was randomised. It is the error of Section~\ref{sec:res-pseudo} moved up one level, and it is invisible in the cross-validation because predicting a held-out plot never requires the strata. We therefore keep R3 as the model of record.

Across ordinal families and fixed-effect structures, the adjacent-category family was worst at both sites (Table~\ref{tab:cv-family}), and the cumulative family ranked first at both. What differs between the sites is the margin, not the direction. At Merredin the sequential family was \MerredinFamSequentialDiff{} behind (SE \MerredinFamSequentialSe{}), which is well inside the noise, while at Wickepin it was \WickepinFamSequentialDiff{} behind (SE \WickepinFamSequentialSe{}), which is not. Section~\ref{sec:res-scale} looks at where that margin comes from. We keep the sequential family as the model of record at both sites, for the mechanistic reasons given in Section~\ref{sec:families}, and report the cumulative fits as a sensitivity check. For treatment structure (Table~\ref{tab:cv-fixed}), the saturated depth $\times$ variety model ranked lowest at both sites, so the data give little predictive support for a depth $\times$ variety--herbicide interaction.

For the model of record (R3) the maximum $\hat{R}$ was \PrimaryMaxRhat{} and the minimum bulk effective sample size \PrimaryMinEss{} over both sites. Across all fitted models the maximum $\hat{R}$ was \MaxRhat{}, and the minimum bulk and tail effective sample sizes were \MinEss{} and \MinEssTail{}. Divergent transitions were rare, \TotalDivergent{} in total over all fits of 24\,000 post-warm-up iterations each, and none occurred under R3 (Table~\ref{tab:diagnostics}). Posterior predictive checks showed that R3 reproduced the observed category frequencies at both sites (Figure~\ref{fig:ppc}).

\subsection{Why the families separate at one site and not the other}
\label{sec:res-scale}

What differs between the sites is not which family wins but whether the two can be told apart at all, and that is the pattern to explain. At Wickepin the cumulative family is ahead of the sequential by \WickepinScaleObsSratioDiff{} (SE \WickepinScaleObsSratioSe{}) on the observed Zadoks codes and by \WickepinScalePrinSratioDiff{} (SE \WickepinScalePrinSratioSe{}) on the principal-stage scale, a large margin that is stable across scales. At Merredin the two sit within three elpd of each other on either scale (\MerredinScaleObsSratioDiff{}, SE \MerredinScaleObsSratioSe{} on the observed codes, and \MerredinScalePrinCumulDiff{}, SE \MerredinScalePrinCumulSe{} in the other direction on the principal stages), so which one is nominally in front changes with the category scale and carries no weight. The model of record stayed sequential at both sites. Our first explanation for the Wickepin margin was the category set rather than the families themselves.

Wickepin recorded three Zadoks codes exactly once each in 900 plants (Z15, Z23 and Z24); Merredin recorded two such codes (Z01 and Z22). The sequential family gives every transition its own unconstrained threshold, which is what lets it handle the leaf-to-tillering discontinuity, but it also means a transition seen once contributes a free parameter estimated from one plant. The cumulative family's thresholds have to be ordered on a shared latent axis, so a sparse category borrows strength from its neighbours, and the ordering constraint costs flexibility while returning some regularisation. Wickepin has more singleton categories than Merredin, so on this account Wickepin should penalise the sequential family more heavily, which is what happens.

The explanation was testable, and the refit did not support it. We refitted both sites on a principal-stage scale, defined on biological grounds before we looked at the outcome: germination (G, Zadoks 00--10), one leaf (Z11), two leaves (Z12), three leaves (Z13), four or more leaves (Z14+), and tillering (T, Zadoks 21 and beyond). Merredin never reached Z14+ and Wickepin never recorded G, so each site loses a different one of these six categories, leaving \MerredinScalePrinK{} categories at each site with none below 19 observations, and putting both sites on a common scale. Table~\ref{tab:family-scale} shows that removing the singleton categories left the Wickepin margin essentially untouched, with the sequential family staying \WickepinScalePrinSratioDiff{} (SE \WickepinScalePrinSratioSe{}) behind cumulative on the principal-stage scale, against \WickepinScaleObsSratioDiff{} on the observed codes. Whatever produces that margin, it is not the three plants that define a category on their own.

\begin{figure}[htbp]
\centering
\includegraphics[width=\textwidth]{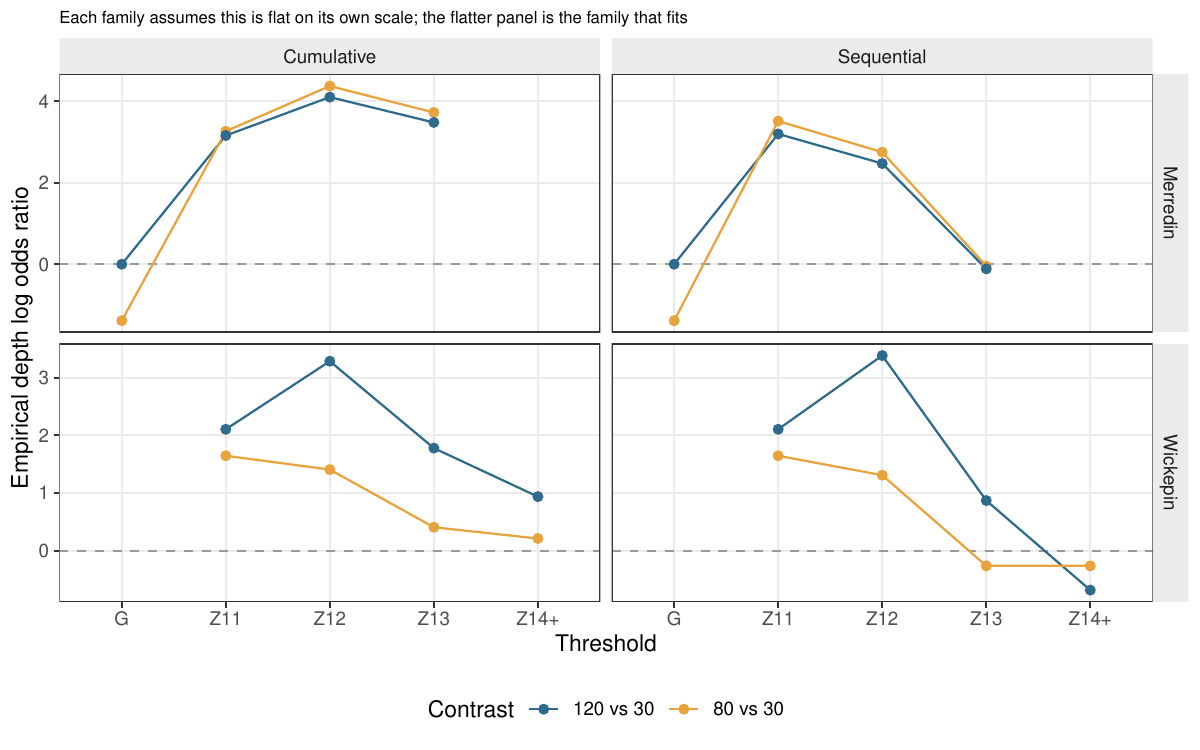}
\caption{Empirical sowing depth log odds ratio at each threshold, computed on the cumulative scale (log odds of $P(Y>k)$) and on the sequential scale (log odds of $P(Y>k \mid Y \geq k)$), on the principal-stage scale. Each family assumes a single treatment coefficient, and so a flat profile on its own scale. The flatter panel points to the family that fits, and it is not the same family at the two sites.}
\label{fig:familyshape}
\end{figure}

We find a better explanation in the shape of the treatment effect across categories, which can be read straight off the contingency tables. Both families use one treatment coefficient, but they hold it constant on different scales: cumulative on the log odds of $P(Y>k)$, sequential on the log odds of $P(Y>k \mid Y \geq k)$. The family whose scale gives the flatter empirical profile is then the one that fits better. Table~\ref{tab:family-shape} and Figure~\ref{fig:familyshape} make the comparison.

The diagnostic requires no model to be fitted, and it anticipates the cross-validation in every site and scale combination we examined. At Wickepin the depth effect is flatter on the cumulative scale on both category scales, and cumulative wins both times. At Merredin the two scales are nearly equally flat, the ordering between them switches when the categories are collapsed, and cross-validation switches with it, by margins small enough to be noise in both cases. The sizes correspond as well as the directions. The two scales differ most at Wickepin, where the predictive gap is around 36 elpd, and least at Merredin, where it is under three. Read biologically, depth at Wickepin shifts developmental position fairly uniformly, so a common shift of every cumulative split describes it well; at Merredin the effect sits mostly in the earlier transitions, and neither proportional form has much advantage over the other.

What is context dependent here is not the mechanism but the proportionality assumption. Development at Wickepin did not switch to some other process; the depth effect there is simply less nearly constant on the sequential scale than on the cumulative one, while at Merredin the data do not distinguish the two. We keep the sequential family at both sites, since it is the one that matches how a plant moves through the scale, and read the comparison as a diagnostic of how hard the single-coefficient assumption is working. On that reading the Wickepin result points to a partially non-proportional sequential model as the natural refinement, rather than to a change of family.

\subsection{The cost of omitting the plot term}
\label{sec:res-pseudo}

\begin{figure}[htbp]
\centering
\includegraphics[width=\textwidth]{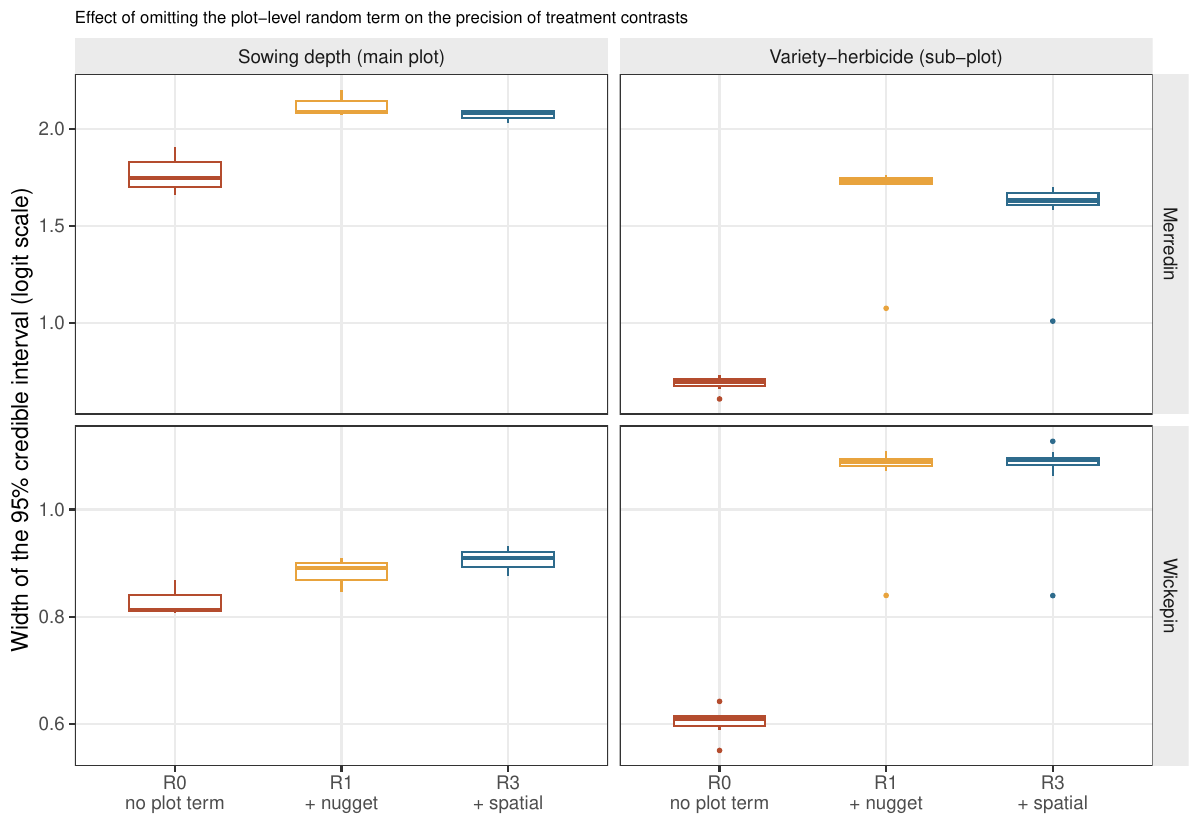}
\caption{Width of the 95\% credible interval for treatment contrasts on the latent logit scale under three random-effect structures. Dropping the plot-level term (R0) barely affects the main-plot factor, which keeps its own error stratum, but it badly overstates the precision of the sub-plot factor, whose error stratum is the plot itself.}
\label{fig:pseudorep}
\end{figure}

We predicted in Section~\ref{sec:units} that leaving out the plot-level random effect would inflate the precision of sub-plot contrasts while leaving main-plot contrasts more or less intact. Table~\ref{tab:pseudoreplication} and Figure~\ref{fig:pseudorep} compare the width of the 95\% credible intervals under R0, R1 and R3, and confirm it.

\begin{table}[htbp]
\centering
\caption{Mean width of the 95\% credible interval for treatment contrasts on the latent logit scale, under three random-effect structures. R0 omits the plot-level term, so the ten plants scored in a plot are treated as ten independent replicates. The main-plot factor (sowing depth) is largely protected because it retains its own error stratum; the sub-plot factor (variety--herbicide), which must be judged against plot-level error, is where omitting the term inflates precision.}
\label{tab:pseudoreplication}
\resizebox{\textwidth}{!}{%
\begin{tabular}{llrrr}
\toprule
Site & Factor & R0: no plot-level term & R1: + plot nugget & R3: + AR1$\times$AR1 + nugget \\
\midrule
Merredin & Sowing depth (main plot) & 1.77 & 2.12 & 2.07 \\
Merredin & Variety-herbicide (sub-plot) & 0.69 & 1.67 & 1.58 \\
Wickepin & Sowing depth (main plot) & 0.83 & 0.88 & 0.91 \\
Wickepin & Variety-herbicide (sub-plot) & 0.60 & 1.06 & 1.07 \\
\bottomrule
\end{tabular}%
}
\end{table}

For the main-plot factor the effect is modest. Mean interval width for the sowing depth contrasts was \MerredinPseudoWidthDepthRZero{} under R0 and \MerredinPseudoWidthDepthRThree{} under R3 at Merredin (a factor of \MerredinPseudoRatioDepth{}), and \WickepinPseudoWidthDepthRZero{} versus \WickepinPseudoWidthDepthRThree{} at Wickepin (\WickepinPseudoRatioDepth{}). Depth retains its own main-plot error stratum in all three models, so it is protected.

For the sub-plot factor the effect is large. Mean interval width for the variety--herbicide effects was \MerredinPseudoWidthVarietyRZero{} under R0 against \MerredinPseudoWidthVarietyRThree{} under R3 at Merredin, a factor of \MerredinPseudoRatioVariety{}, and \WickepinPseudoWidthVarietyRZero{} against \WickepinPseudoWidthVarietyRThree{} at Wickepin (\WickepinPseudoRatioVariety{}).

The size of the effect is about what we would expect from subsampling. Taking the design effect for a mean as a rough benchmark, with $n = 10$ observational units per experimental unit the inflation factor is $\sqrt{1 + (n-1)\rho}$ for within-plot correlation $\rho$, so the often-quoted $\sqrt{10} \approx 3.2$ is an upper bound, reached only if plants in a plot were perfectly correlated. The ratios here, \MerredinPseudoRatioVariety{} at Merredin and \WickepinPseudoRatioVariety{} at Wickepin, correspond to moderate within-plot correlation, and the larger value at Merredin goes with the larger plot-level variance found there (Section~\ref{sec:varcomp}). A model without the plot term would therefore report variety--herbicide differences as well resolved when the design does not resolve them. The error is invisible in the fit statistics of that model on its own. It shows up only against a model that carries the stratum, or under cross-validation that holds out whole plots.

\subsection{Variance components and spatial structure}
\label{sec:varcomp}

\begin{figure}[htbp]
\centering
\includegraphics[width=\textwidth]{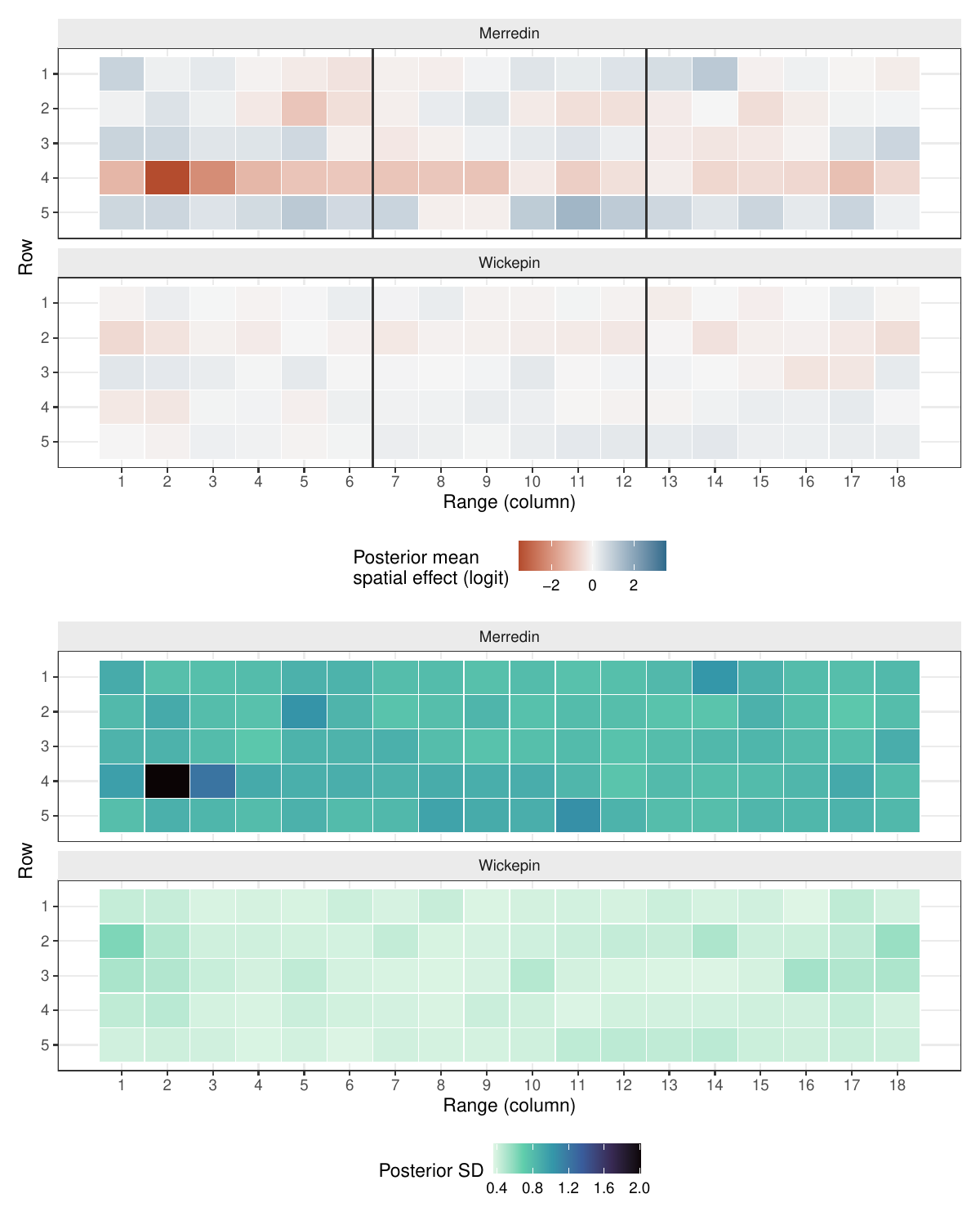}
\caption{Posterior mean (top) and posterior standard deviation (bottom) of the fitted AR1\,$\times$\,AR1 spatial field on the latent logit scale. Grey lines mark the replicate boundaries.}
\label{fig:field}
\end{figure}

Table~\ref{tab:varcomp} gives the posterior summaries for the model of record at each site, and Figure~\ref{fig:field} maps the posterior mean and standard deviation of the fitted spatial field.

\begin{table}[htbp]
\centering
\caption{Posterior mean and 95\% credible interval for the variance components and autocorrelation parameters of the primary model (sequential family, depth $\times$ variety fixed effects, replicate and main-plot random effects, AR1$\times$AR1 spatial field and plot nugget). All quantities are on the latent logit scale.}
\label{tab:varcomp}
\small
\begin{tabular}{lll}
\toprule
Parameter & Merredin & Wickepin \\
\midrule
$\sigma_{\mathrm{rep}}$ (replicate) & 0.58 [0.01, 2.09] & 0.29 [0.01, 1.15] \\
$\sigma_{\mathrm{mp}}$ (main plot: rep $\times$ depth) & 1.33 [0.29, 2.65] & 0.33 [0.02, 0.88] \\
$\sigma_s$ (AR1$\times$AR1 field) & 1.09 [0.25, 1.79] & 0.41 [0.02, 0.89] \\
$\sigma_e$ (plot nugget) & 0.81 [0.04, 1.46] & 0.58 [0.07, 0.91] \\
$\rho_r$ (autocorrelation along rows) & -0.11 [-0.76, 0.51] & -0.04 [-0.74, 0.72] \\
$\rho_c$ (autocorrelation along ranges) & 0.58 [-0.05, 0.96] & 0.28 [-0.65, 0.92] \\
\bottomrule
\end{tabular}
\end{table}

The two sites rank their variance components differently. At Merredin the main-plot stratum is the largest ($\sigma_{\mathrm{mp}} = \MerredinSdWp{}$, 95\% CI \MerredinSdWpCI{}), followed by the spatial field ($\sigma_s = \MerredinSdSpatial{}$, \MerredinSdSpatialCI{}) and then the independent plot nugget ($\sigma_e = \MerredinSdNugget{}$, \MerredinSdNuggetCI{}). At Wickepin the order reverses, with the nugget largest ($\sigma_e = \WickepinSdNugget{}$, \WickepinSdNuggetCI{}), ahead of the spatial field ($\sigma_s = \WickepinSdSpatial{}$, \WickepinSdSpatialCI{}) and the main-plot term ($\sigma_{\mathrm{mp}} = \WickepinSdWp{}$, \WickepinSdWpCI{}). The replicate standard deviation was the smallest component at both sites. Structured variation (main-plot and spatial) therefore dominates at Merredin, while unstructured plot-to-plot variation dominates at Wickepin. The cross-validation does not mirror that contrast. A variance component can be sizeable without the correlation structure that carries it being recoverable from 90 plots on a 5-row grid, and it is the structure, not the size, that out-of-sample prediction is sensitive to.

The autocorrelation parameters were weakly identified, as our simulation study had anticipated. At Merredin the posterior for the along-row autocorrelation was $\rho_r = \MerredinArRow{}$ (\MerredinArRowCI{}) and along ranges $\rho_c = \MerredinArCol{}$ (\MerredinArColCI{}); at Wickepin $\rho_r = \WickepinArRow{}$ (\WickepinArRowCI{}) and $\rho_c = \WickepinArCol{}$ (\WickepinArColCI{}). The credible intervals cover most of the permitted range in every case. This is a property of the design rather than of the model. With five rows there is very little information about $\rho_r$, and along ranges the replicate and main-plot terms are themselves contiguous strips, so they already absorb most of the coherent variation an autoregressive process could explain. The fitted fields in Figure~\ref{fig:field} are correspondingly smooth and small relative to the nugget.

\subsection{Sowing depth effects}
\label{sec:depth}

\begin{figure}[htbp]
\centering
\includegraphics[width=\textwidth]{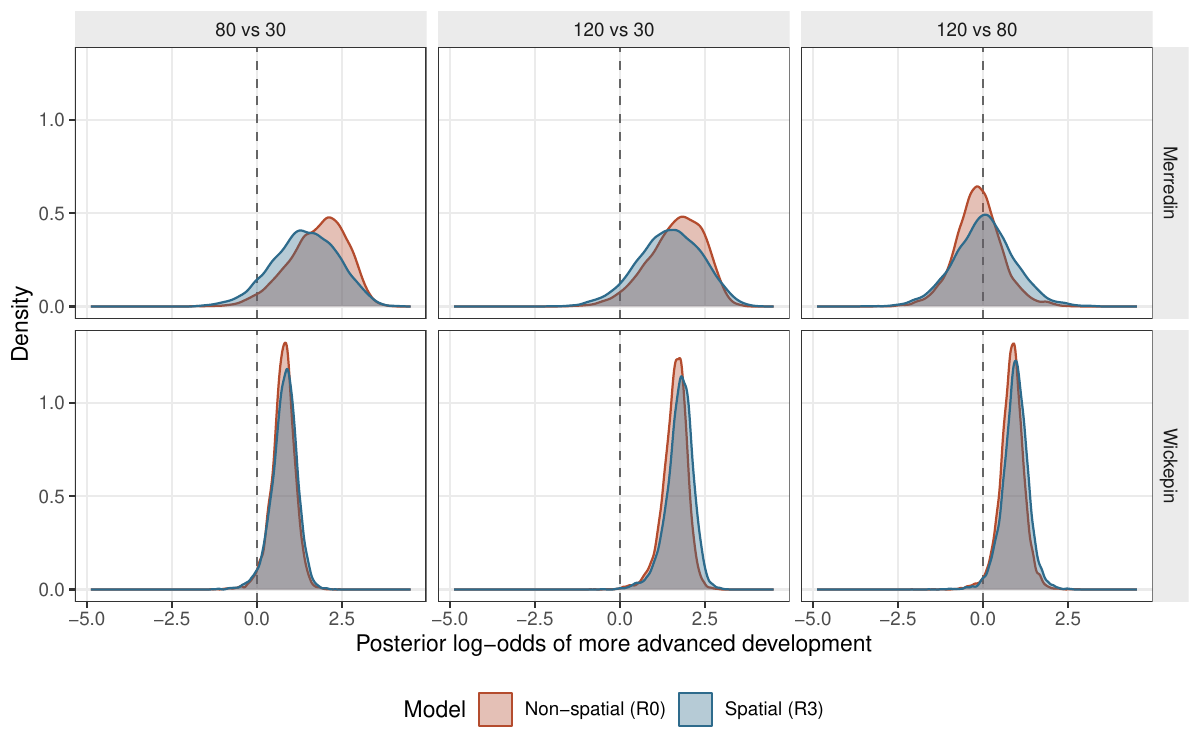}
\caption{Posterior distributions of the variety-averaged sowing depth contrasts on the latent logit scale, with and without the AR1\,$\times$\,AR1 spatial field. Positive values indicate that the deeper treatment is more advanced. The two sets of posteriors overlap closely, which is the main robustness result of the paper.}
\label{fig:contrast}
\end{figure}

Table~\ref{tab:depth-contrasts} and Figure~\ref{fig:contrast} give the variety-averaged depth contrasts on the latent logit scale, fitted with and without the spatial term.

\begin{table}[htbp]
\centering
\caption{Variety-averaged sowing depth contrasts on the latent logit scale, with and without the AR1$\times$AR1 spatial field. A positive value means the deeper treatment is developmentally more advanced four weeks after sowing. $P(\text{effect}>0)$ is the posterior probability that the contrast is positive.}
\label{tab:depth-contrasts}
\small
\begin{tabular}{lllrr}
\toprule
Site & Model & Depth contrast (mm) & Log-odds [95\% CI] & P(effect $>0$) \\
\midrule
Merredin & Spatial (R3) & 80 vs 30 & 1.42 [-0.51, 3.13] & 0.928 \\
Merredin & Non-spatial (R0) & 80 vs 30 & 1.78 [-0.07, 3.16] & 0.971 \\
Merredin & Spatial (R3) & 120 vs 30 & 1.46 [-0.47, 3.13] & 0.936 \\
Merredin & Non-spatial (R0) & 120 vs 30 & 1.65 [-0.13, 2.97] & 0.965 \\
Merredin & Spatial (R3) & 120 vs 80 & 0.04 [-1.75, 1.79] & 0.519 \\
Merredin & Non-spatial (R0) & 120 vs 80 & -0.13 [-1.54, 1.34] & 0.408 \\
Wickepin & Spatial (R3) & 80 vs 30 & 0.80 [-0.04, 1.48] & 0.970 \\
Wickepin & Non-spatial (R0) & 80 vs 30 & 0.77 [0.02, 1.40] & 0.977 \\
Wickepin & Spatial (R3) & 120 vs 30 & 1.75 [0.83, 2.43] & 0.999 \\
Wickepin & Non-spatial (R0) & 120 vs 30 & 1.61 [0.76, 2.24] & 0.999 \\
Wickepin & Spatial (R3) & 120 vs 80 & 0.95 [0.16, 1.69] & 0.985 \\
Wickepin & Non-spatial (R0) & 120 vs 80 & 0.85 [0.10, 1.52] & 0.982 \\
\bottomrule
\end{tabular}
\end{table}

Positive values indicate that the deeper treatment is developmentally more advanced four weeks after sowing.

At Merredin, sowing at 80~mm rather than 30~mm gave a contrast of \MerredinDepthEightyVsThirty{} (95\% CI \MerredinDepthEightyVsThirtyCI{}; posterior probability of a positive effect \MerredinDepthEightyVsThirtyP{}), and sowing at 120~mm rather than 30~mm gave \MerredinDepthOneTwentyVsThirty{} (95\% CI \MerredinDepthOneTwentyVsThirtyCI{}; $P = \MerredinDepthOneTwentyVsThirtyP{}$). The additional 40~mm from 80 to 120~mm produced little further change (\MerredinDepthOneTwentyVsEighty{}, 95\% CI \MerredinDepthOneTwentyVsEightyCI{}).

At Wickepin the pattern was the same and graded, with \WickepinDepthEightyVsThirty{} (95\% CI \WickepinDepthEightyVsThirtyCI{}; $P = \WickepinDepthEightyVsThirtyP{}$) for 80 versus 30~mm and \WickepinDepthOneTwentyVsThirty{} (95\% CI \WickepinDepthOneTwentyVsThirtyCI{}; $P = \WickepinDepthOneTwentyVsThirtyP{}$) for 120 versus 30~mm, and a further increment of \WickepinDepthOneTwentyVsEighty{} (95\% CI \WickepinDepthOneTwentyVsEightyCI{}) from 80 to 120~mm.

Fitting the spatial field changed these conclusions very little. Under the non-spatial split-plot model the corresponding 120 versus 30~mm contrasts were \MerredinDepthNSOneTwentyVsThirty{} (95\% CI \MerredinDepthNSOneTwentyVsThirtyCI{}) at Merredin and \WickepinDepthNSOneTwentyVsThirty{} (95\% CI \WickepinDepthNSOneTwentyVsThirtyCI{}) at Wickepin. Figure~\ref{fig:contrast} shows the two sets of posteriors sitting almost on top of each other at both sites. This matters because the depth strips are contiguous and could in principle have been confounded with a field trend. The depth effect persists when the spatial structure that might have produced it artefactually is modelled.

Variety--herbicide effects were less clearly resolved than depth (Table~\ref{tab:variety}). Each combination, averaged over sowing depth, differed from the average combination by at most \MerredinVhMaxAbs{} logits at Merredin and \WickepinVhMaxAbs{} at Wickepin, against 120 versus 30~mm depth contrasts of \MerredinDepthOneTwentyVsThirty{} and \WickepinDepthOneTwentyVsThirty{}, and the 95\% credible interval excluded zero for only \MerredinVhNExclZero{} of the 10 combinations at Merredin and \WickepinVhNExclZero{} at Wickepin.

\subsection{Prior sensitivity}
\label{sec:sensitivity}

The recovery study showed the $\mathcal{N}(0,1)$ prior shrinking large contrasts and the Beta$(2,2)$ prior pulling the autocorrelations toward zero, so we refitted the model of record with each relaxed (Table~\ref{tab:sensitivity}). Widening the contrast prior to $\mathcal{N}(0, 2.5)$ moved the 120 versus 30~mm effect from \MerredinSensPrimary{} to \MerredinSensWideBeta{} at Merredin and from \WickepinSensPrimary{} to \WickepinSensWideBeta{} at Wickepin, leaving the direction and the practical conclusion unchanged; the primary analysis is the more conservative of the two. Flattening the autocorrelation prior to uniform on $(-1,1)$ left the depth contrasts essentially untouched (\MerredinSensFlatRho{} and \WickepinSensFlatRho{}) while allowing $\rho_c$ to move to \MerredinSensFlatRhoCol{} and \WickepinSensFlatRhoCol{}. The autocorrelations respond to their prior and the treatment effects do not, and that contrast sums up the spatial analysis as a whole.

\subsection{Predicted stage probabilities and tillering}
\label{sec:probs}

Table~\ref{tab:pred-probs} and Figure~\ref{fig:predprobs} give the posterior predictive probability that a plant in a new plot is at each Zadoks stage four weeks after sowing, averaged over the ten variety--herbicide combinations. These predictions integrate over the replicate, main-plot, spatial and plot-level random effects, so they describe a plot the model has not seen rather than an idealised plot with all random effects set to zero. That is the quantity relevant to a grower planning next season.

\begin{figure}[htbp]
\centering
\includegraphics[width=0.75\textwidth]{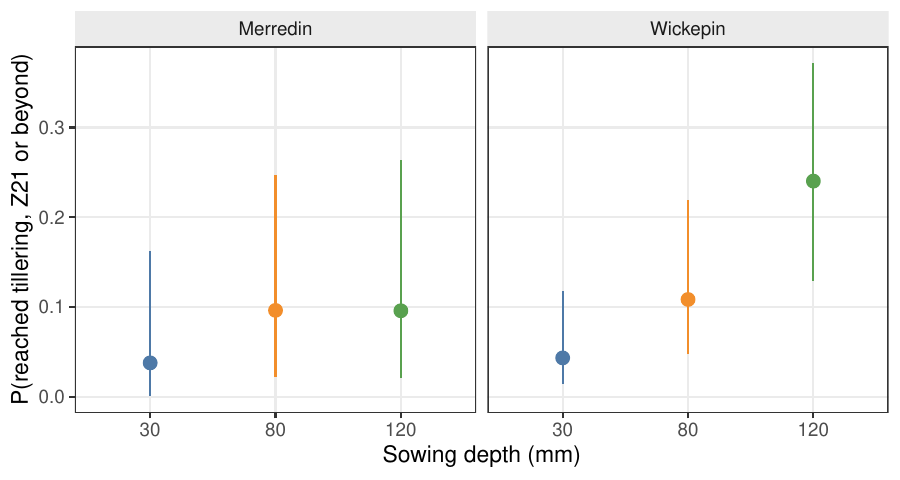}
\caption{Posterior probability that a plant in a new plot has reached tillering (Z21 or beyond) four weeks after sowing, by site and sowing depth.}
\label{fig:tillering}
\end{figure}

The probability of having reached tillering (Z21 or beyond), Figure~\ref{fig:tillering}, rose with depth at both sites.

At Merredin it was \MerredinTillThirty{} (95\% CI \MerredinTillThirtyCI{}) at 30~mm, \MerredinTillEighty{} (\MerredinTillEightyCI{}) at 80~mm and \MerredinTillOneTwenty{} (\MerredinTillOneTwentyCI{}) at 120~mm. At Wickepin the corresponding probabilities were \WickepinTillThirty{} (\WickepinTillThirtyCI{}), \WickepinTillEighty{} (\WickepinTillEightyCI{}) and \WickepinTillOneTwenty{} (\WickepinTillOneTwentyCI{}). The posterior probability that sowing at 120~mm raises the chance of tillering relative to 30~mm was \MerredinTillPOneTwentyVsThirty{} at Merredin and \WickepinTillPOneTwentyVsThirty{} at Wickepin.

\subsection{Threshold structure}
\label{sec:threshold}

The sequential model estimates one parameter per transition (Table~\ref{tab:thresholds}, Figure~\ref{fig:thresholds}), but with a single census those parameters cannot be read as ``transition difficulty''. The profile mostly reflects how far a progressing population had got by the observation date, and sparse categories make individual estimates unstable as well (the singleton stages at Wickepin, for example). We therefore treat the thresholds as a flexible way of representing the scale, not as biological parameters \citep{Slafer1994}.

\subsection{Development as a discrete-time hazard}
\label{sec:res-hazard}

Using a complementary log-log link recasts the sequential model as grouped-time proportional hazards \citep{Prentice1978, Laara1985}. We state explicitly which event carries the proportional hazard, because it is not the one the name suggests. Writing $h_k = P(Y_i > k \mid Y_i \geq k)$ for the conditional probability of advancing, the link gives $-\log h_k = e^{-\alpha_k} e^{-\eta_i}$, so for the complementary event, the plant halting at stage $k$,
\begin{equation}
  \mathrm{cloglog}\bigl(P(Y_i = k \mid Y_i \geq k)\bigr) = -\alpha_k - \eta_i .
  \label{eq:cloglog}
\end{equation}
The hazard is the hazard of \emph{halting}, its linear predictor is $-\eta_i$, and the hazard ratio is $\exp(-\beta)$. A value below one means the deeper treatment is less likely to stall at any given transition, and so is developmentally more advanced. (Section~\ref{sec:families} used ``hazard'' descriptively for the conditional advancement probability of Equation~\eqref{eq:sratio}; a proportional hazards model for advancing would be the continuation-ratio complementary log-log form, which is a different model and is not what is fitted here.)

This form did not improve fit. It was comparable at Merredin (\MerredinLinkDiff{}, SE \MerredinLinkSe{}) and clearly worse at Wickepin (\WickepinLinkDiff{}, SE \WickepinLinkSe{}). We therefore treat it as an interpretive alternative, not a preferred model. Read on that scale (Table~\ref{tab:hazard}, Figure~\ref{fig:hazard}), sowing at 120 rather than 30~mm multiplied the hazard of halting by \MerredinHROneTwentyVsThirty{} at Merredin and \WickepinHROneTwentyVsThirty{} at Wickepin, which is the conclusion Section~\ref{sec:depth} reached on the logit scale, expressed as a rate of developmental progress rather than as a shift in position.

\subsection{Cross-site consistency}
\label{sec:crosssite}

The direction of the depth effect was the same at both sites, with deeper sowing advancing development at four weeks in both trials (Figure~\ref{fig:contrast}). Pairing the two sites' independently fitted posteriors, the 80 versus 30~mm and 120 versus 30~mm contrasts had the same sign at both sites with posterior probability \CrossSiteEightyVsThirtySameSign{} and \CrossSiteOneTwentyVsThirtySameSign{} respectively (Table~\ref{tab:cross-site}). The shape of the response was not. Wickepin gave a graded pattern, while Merredin looked closer to a step change between 30 and 80~mm, and the 120 versus 80~mm increment agreed in sign across sites with probability only \CrossSiteOneTwentyVsEightySameSign{}, which is no better than chance. In practice the direction of the effect generalises more reliably than its size or its dose response form \citep{Piepho2008met}.

\section{Discussion}
\label{sec:discussion}

\subsection{Deep sowing advanced early development at both sites}

Deeper sowing went with more advanced development at four weeks at both sites. Deeper placement can delay emergence mechanically, but early establishment in these environments appears to be limited by moisture, so getting the seed into stored subsoil water probably outweighed the extra time needed for coleoptile extension \citep{Rebetzke2007, Kirkegaard2014, Fletcher2015}. At Merredin the gain beyond 80~mm was small and came with more plants at very early stages, which suggests a local trade-off between advancement and establishment risk.

\subsection{What the spatial model contributed}
\label{sec:disc-spatial}

Three things came out of the spatial model, and only the first is the one we expected. Plot-level heterogeneity was essential, both for prediction and for getting the uncertainty right, and it accounted for almost the whole gap between the classical split-plot model and the best model at each site. Letting that heterogeneity be spatially correlated then added nothing at either site. On a 5\,$\times$\,18 grid with contiguous replicates this is not surprising. Five rows carry very little information about $\rho_r$, and the replicate and main-plot strips already absorb most of the coherent variation an autoregressive process could explain. It is still worth stating plainly, because the usual argument for AR1\,$\times$\,AR1 in field trials is made on layouts with far more columns \citep{Cullis1991, Gilmour1997, Stefanova2009}.

The third is a caution rather than a finding. Dropping the blocking terms entirely (R4) predicted a held-out plot as well as keeping them at both sites. It also narrowed the intervals for the sowing depth contrasts, to \MerredinDepthWidthRatioRFour{} of their R3 width at Merredin and \WickepinDepthWidthRatioRFour{} at Wickepin, and at Merredin it turned an interval that covers zero into one that does not (Section~\ref{sec:modelcomp-results}). Predicting a plot and estimating a main-plot contrast are different questions, and only the second needs the stratum against which the factor was randomised. Cross-validation cannot see the difference, so it should not be the sole arbiter of a random-effect structure when the target of inference is a treatment effect. The point concerns which question cross-validation answers, not whether it should be used. On this layout, then, the spatial field is best read as a robustness check on the treatment conclusions, and not as an estimate of a spatial process.

\subsection{Subsampling changes both the model and the cross-validation}
\label{sec:disc-units}

Subsampling creates two linked requirements. On the modelling side, plants are observational units and plots are experimental units, so a plant-level ordinal analysis needs a plot-level random effect. Without it the uncertainty is understated, particularly for sub-plot factors (Section~\ref{sec:res-pseudo}) \citep{Hurlbert1984, Piepho2003, Bolker2009}. On the evaluation side, the held-out unit should be the plot as well, since leave-one-plant-out mostly rewards interpolation inside a plot the model has already seen.

The same logic explains why PSIS-LOO was unstable here. Leaving out a plot strongly perturbs the posterior for that plot's own latent terms, which gives heavy-tailed importance weights and large Pareto $\hat{k}$ values. Exact $K$-fold over plots is the more reliable basis for comparing random-effect and spatial structures.

\subsection{What the family comparison is, and is not, evidence about}
\label{sec:disc-family}

Cross-validation ruled out the adjacent-category family at both sites. It separated cumulative from sequential only at Wickepin; at Merredin the two were within noise on either category scale. Sparse categories did not explain the Wickepin margin (Section~\ref{sec:res-scale}). What the margin tracked was which scale carried the more nearly constant treatment effect, and a diagnostic read straight off the contingency tables, with no model fitted, anticipated both its direction and its size. Predictive accuracy compares proportional forms, and a family loses when its own constancy assumption is strained, which does not necessarily imply that the underlying developmental biology changed. Selecting family purely by empirical fit per site would place the two sites' coefficients on different scales, complicating cross-site comparison (Section~\ref{sec:crosssite}). We retained the sequential family as the model of record because it matches the irreversible, stepwise way in which a cereal plant passes through growth stages, and used the cross-validation and the empirical diagnostic as checks on the single-coefficient assumption.

Substantive inference, on the other hand, was stable across families. The depth effects kept the same direction and much the same size, and the predicted tillering probabilities were close under the cumulative and sequential fits. In this data set the family mattered for efficiency and interpretation, not for the agronomic conclusion.

\subsection{Limitations}

Stage was measured at a single census date four weeks after sowing, so the separate dynamics of emergence timing versus subsequent developmental rate cannot be disentangled. Some intermediate categories were sparsely observed. The data also come from two sites in a single season, which limits what can be said about the stability of the depth effect across years. Finally, the Merredin scale maps germination and leaf stages onto a single ordinal continuum, which is practical for modelling but biologically coarser than a joint bivariate model of emergence and vegetative growth.

\subsection{Broader applicability and extension}

The framework carries over to other ordinal traits scored on plot grids, such as disease severity, lodging and maturity class, where subsampling and spatial structure are often handled badly. One extension is a multi-environment hierarchical model that separates the consistent part of the depth effect from the environment-specific part, and reports the posterior probability that a recommendation holds across the target environments \citep{Piepho2008met}. Another is to relax the assumption that the treatment effect is constant across the field. Here depth enters as a single coefficient and all spatial variation is pushed into the error, but for continuous responses the treatment parameters can be given their own spatial process \citep{cao2022bayesian}. On an ordinal scale the same idea would ask whether deep sowing helps more in the parts of a paddock that hold more subsoil moisture, a question this design cannot answer but a grower would want answered.

\section{Conclusions}
\label{sec:conclusions}

We specified, validated and applied a Bayesian hierarchical ordinal model with split-plot random effects and an AR1\,$\times$\,AR1 spatial field to growth-stage data from two deep-sowing oat trials.

\begin{itemize}
\item Getting the experimental unit right mattered. A plot-level random effect changed the uncertainty of the sub-plot contrasts substantially, and models without it looked more precise than they were.
\item For subsampled ordinal data, model comparison was most stable under exact $K$-fold cross-validation over plots. What the data asked for was plot-level heterogeneity; giving it an AR1\,$\times$\,AR1 correlation structure added nothing at either site on this 5\,$\times$\,18 layout. Cross-validation could not separate the model without blocking terms from the full one, which is a reason to keep inference on main-plot factors anchored to the design-based strata rather than to predictive fit.
\item Agronomically, deeper sowing (80 and 120~mm against 30~mm) went with more advanced development at four weeks at both sites (the posterior probability of a positive 120 versus 30~mm effect was \MerredinDepthOneTwentyVsThirtyP{} at Merredin and \WickepinDepthOneTwentyVsThirtyP{} at Wickepin), and that held across the spatial and ordinal-family specifications we tried.
\end{itemize}

\section*{Data and code availability}

Data are not available. Code will be made available upon publication.

\section*{Acknowledgements}

This work was supported by GRDC, in collaboration with the Facey Group and Living Farm. The authors thank field staff at Merredin and Wickepin for trial establishment and data collection.

\section*{Author contributions}

Z.C.\ conceived the statistical methodology, implemented and validated the Stan model, performed the analysis, and drafted the manuscript. R.M.\ and M.N.\ conducted data processing, trial design and assisted with visualisation. A.H.S.\ contributed agronomic expertise, and revised the manuscript. All authors read and approved the final manuscript.

\section*{Conflict of interest}

The authors declare no conflict of interest.

\bibliographystyle{elsarticle-harv}
\bibliography{references}

\clearpage
\appendix
\singlespacing
\small
\renewcommand{\thetable}{S\arabic{table}}
\renewcommand{\thefigure}{S\arabic{figure}}
\setcounter{table}{0}
\setcounter{figure}{0}

\section{Supplementary tables and figures}
\label{app:supp}

\begin{table}[htbp]
\centering
\caption{Exact five-fold cross-validation over plots for the three ordinal families, fitted under the primary random-effect structure at each site. Elpd is a log predictive density, so higher (less negative) is better; differences are measured against the best model in the block, and are therefore never positive.}
\label{tab:cv-family}
\small
\begin{tabular}{llrrrr}
\toprule
Site & Model & elpd$_{5\mathrm{-fold}}$ & SE & Difference & SE of difference \\
\midrule
Merredin & Cumulative & -895.5 & 55.3 & 0.0 & 0.0 \\
Merredin & Sequential & -898.4 & 56.9 & -2.9 & 3.5 \\
Merredin & Adjacent-category & -922.7 & 59.3 & -27.2 & 8.9 \\
Wickepin & Cumulative & -1121.5 & 30.9 & 0.0 & 0.0 \\
Wickepin & Sequential & -1158.7 & 33.2 & -37.2 & 8.1 \\
Wickepin & Adjacent-category & -1175.7 & 33.0 & -54.2 & 10.1 \\
\bottomrule
\end{tabular}
\end{table}

\begin{table}[htbp]
\centering
\caption{Exact five-fold cross-validation over plots for the fixed-effect structure, fitted under the primary random-effect structure and the sequential family. Elpd is a log predictive density, so higher (less negative) is better; differences are measured against the best model in the block, and are therefore never positive.}
\label{tab:cv-fixed}
\small
\begin{tabular}{llrrrr}
\toprule
Site & Model & elpd$_{5\mathrm{-fold}}$ & SE & Difference & SE of difference \\
\midrule
Merredin & Depth + variety & -882.9 & 52.0 & 0.0 & 0.0 \\
Merredin & Depth only & -884.8 & 52.9 & -1.9 & 3.4 \\
Merredin & Depth $\times$ variety & -898.4 & 56.9 & -15.5 & 8.1 \\
Wickepin & Depth + variety & -1153.6 & 33.0 & 0.0 & 0.0 \\
Wickepin & Depth only & -1154.2 & 32.6 & -0.6 & 3.3 \\
Wickepin & Depth $\times$ variety & -1158.7 & 33.2 & -5.1 & 3.0 \\
\bottomrule
\end{tabular}
\end{table}

\begin{table}[htbp]
\centering
\caption{Exact five-fold cross-validation over plots for the three ordinal families on two category scales. The observed scale keeps one category per Zadoks code recorded at a site, including codes seen only once; the principal-stage scale collapses these to germination (G), one leaf (Z11), two leaves (Z12), three leaves (Z13), four or more leaves (Z14+), and tillering (T). Merredin never reached Z14+ and Wickepin never recorded G, so each site drops a different one of these six categories, leaving five at each site with none below 19 observations. Differences are measured against the best family within each site and scale.}
\label{tab:family-scale}
\small
\begin{tabular}{lllrrrr}
\toprule
Site & Scale & $K$ & Family & elpd$_{5\mathrm{-fold}}$ & Difference & SE of difference \\
\midrule
Merredin & Observed Zadoks codes & 7 & Cumulative & -895.5 & 0.0 & 0.0 \\
Merredin & Observed Zadoks codes & 7 & Sequential & -898.4 & -2.9 & 3.5 \\
Merredin & Observed Zadoks codes & 7 & Adjacent-category & -922.7 & -27.2 & 8.9 \\
Wickepin & Observed Zadoks codes & 8 & Cumulative & -1121.5 & 0.0 & 0.0 \\
Wickepin & Observed Zadoks codes & 8 & Sequential & -1158.7 & -37.2 & 8.1 \\
Wickepin & Observed Zadoks codes & 8 & Adjacent-category & -1175.7 & -54.2 & 10.1 \\
Merredin & Principal stages & 5 & Sequential & -880.3 & 0.0 & 0.0 \\
Merredin & Principal stages & 5 & Cumulative & -880.3 & -0.1 & 1.9 \\
Merredin & Principal stages & 5 & Adjacent-category & -893.6 & -13.4 & 3.6 \\
Wickepin & Principal stages & 5 & Cumulative & -1100.7 & 0.0 & 0.0 \\
Wickepin & Principal stages & 5 & Adjacent-category & -1125.9 & -25.2 & 6.1 \\
Wickepin & Principal stages & 5 & Sequential & -1136.7 & -36.0 & 7.9 \\
\bottomrule
\end{tabular}
\end{table}

\begin{table}[htbp]
\centering
\caption{Standard deviation, across thresholds, of the empirical depth log odds ratio computed on each family's own scale: the log odds of $P(Y>k)$ for the cumulative family and of $P(Y>k \mid Y \geq k)$ for the sequential family. Each family assumes a common treatment effect on its own scale, so the family whose scale gives the flatter profile is the one whose proportionality assumption the data support.}
\label{tab:family-shape}
\resizebox{\textwidth}{!}{%
\begin{tabular}{lllrrl}
\toprule
Site & Scale & Depth contrast (mm) & SD, cumulative scale & SD, sequential scale & Flatter scale \\
\midrule
Merredin & Observed Zadoks codes & 120 vs 30 & 1.83 & 1.98 & Cumulative \\
Merredin & Observed Zadoks codes & 80 vs 30 & 2.64 & 2.70 & Cumulative \\
Wickepin & Observed Zadoks codes & 120 vs 30 & 0.83 & 1.43 & Cumulative \\
Wickepin & Observed Zadoks codes & 80 vs 30 & 0.68 & 0.80 & Cumulative \\
Merredin & Principal stages & 120 vs 30 & 1.83 & 1.70 & Sequential \\
Merredin & Principal stages & 80 vs 30 & 2.62 & 2.31 & Sequential \\
Wickepin & Principal stages & 120 vs 30 & 0.97 & 1.74 & Cumulative \\
Wickepin & Principal stages & 80 vs 30 & 0.71 & 1.01 & Cumulative \\
\bottomrule
\end{tabular}%
}
\end{table}

\begin{table}[htbp]
\centering
\caption{Pareto $k$ diagnostics for approximate leave-one-plot-out cross-validation by Pareto-smoothed importance sampling, computed from the same Stage~1 fits reported in Table~\ref{tab:cv-random} (sequential family, saturated depth $\times$ variety fixed effects). Values above 0.7 indicate that the importance-sampling approximation is unreliable. The models carrying plot-level random terms fail this diagnostic for a large fraction of plots, which is why the model comparison in this paper uses exact five-fold cross-validation instead.}
\label{tab:pareto-k}
\small
\begin{tabular}{llrrr}
\toprule
Site & Model & Plots with $k>0.7$ & Plots with $k>1$ & max $k$ \\
\midrule
Merredin & R0: rep + main-plot & 16 & 3 & 1.63 \\
Merredin & R1: R0 + plot nugget & 80 & 12 & 1.57 \\
Merredin & R2: R0 + AR1$\times$AR1 & 70 & 10 & 1.5 \\
Merredin & R3: R0 + AR1$\times$AR1 + nugget & 78 & 9 & 1.32 \\
Merredin & R4: AR1$\times$AR1 + nugget, no blocking & 75 & 6 & 1.36 \\
Wickepin & R0: rep + main-plot & 15 & 1 & 1.07 \\
Wickepin & R1: R0 + plot nugget & 73 & 6 & 1.16 \\
Wickepin & R2: R0 + AR1$\times$AR1 & 73 & 5 & 1.13 \\
Wickepin & R3: R0 + AR1$\times$AR1 + nugget & 67 & 7 & 1.16 \\
Wickepin & R4: AR1$\times$AR1 + nugget, no blocking & 66 & 4 & 1.19 \\
\bottomrule
\end{tabular}
\end{table}

\begin{table}[htbp]
\centering
\caption{Robustness of the variety-averaged sowing depth contrasts and of the posterior mean autocorrelations. The primary analysis uses the saturated depth $\times$ variety structure with $\beta \sim \mathcal{N}(0,1)$ and $(\rho+1)/2 \sim \mathrm{Beta}(2,2)$. The alternatives widen the prior on the treatment contrasts, flatten the prior on the autocorrelations, and drop the depth $\times$ variety interaction respectively; all use the same random-effect and spatial structure.}
\label{tab:sensitivity}
\small
\begin{tabular}{lllrr}
\toprule
Site & Prior & Depth contrast (mm) & Log-odds [95\% CI] & $\rho_r$ / $\rho_c$ \\
\midrule
Merredin & Additive depth + variety & 80 vs 30 & 1.32 [-0.52, 2.98] & -0.16 / 0.53 \\
Merredin & Flat: $(\rho+1)/2 \sim$ Beta$(1,1)$ & 80 vs 30 & 1.40 [-0.55, 3.15] & -0.26 / 0.78 \\
Merredin & Primary (saturated, N(0,1)) & 80 vs 30 & 1.42 [-0.51, 3.13] & -0.11 / 0.58 \\
Merredin & Wider: $\beta \sim \mathcal{N}(0, 2.5)$ & 80 vs 30 & 3.59 [1.99, 4.83] & -0.10 / 0.58 \\
Merredin & Additive depth + variety & 120 vs 30 & 1.31 [-0.59, 2.98] & -0.16 / 0.53 \\
Merredin & Flat: $(\rho+1)/2 \sim$ Beta$(1,1)$ & 120 vs 30 & 1.42 [-0.46, 3.11] & -0.26 / 0.78 \\
Merredin & Primary (saturated, N(0,1)) & 120 vs 30 & 1.46 [-0.47, 3.13] & -0.11 / 0.58 \\
Merredin & Wider: $\beta \sim \mathcal{N}(0, 2.5)$ & 120 vs 30 & 3.54 [1.97, 4.74] & -0.10 / 0.58 \\
Merredin & Additive depth + variety & 120 vs 80 & -0.02 [-1.80, 1.79] & -0.16 / 0.53 \\
Merredin & Flat: $(\rho+1)/2 \sim$ Beta$(1,1)$ & 120 vs 80 & 0.02 [-1.82, 1.87] & -0.26 / 0.78 \\
Merredin & Primary (saturated, N(0,1)) & 120 vs 80 & 0.04 [-1.75, 1.79] & -0.11 / 0.58 \\
Merredin & Wider: $\beta \sim \mathcal{N}(0, 2.5)$ & 120 vs 80 & -0.04 [-1.30, 1.24] & -0.10 / 0.58 \\
Wickepin & Additive depth + variety & 80 vs 30 & 0.78 [-0.06, 1.45] & 0.08 / 0.36 \\
Wickepin & Flat: $(\rho+1)/2 \sim$ Beta$(1,1)$ & 80 vs 30 & 0.81 [-0.08, 1.50] & 0.00 / 0.48 \\
Wickepin & Primary (saturated, N(0,1)) & 80 vs 30 & 0.80 [-0.04, 1.48] & -0.04 / 0.28 \\
Wickepin & Wider: $\beta \sim \mathcal{N}(0, 2.5)$ & 80 vs 30 & 1.01 [0.20, 1.77] & -0.16 / 0.23 \\
Wickepin & Additive depth + variety & 120 vs 30 & 1.72 [0.80, 2.39] & 0.08 / 0.36 \\
Wickepin & Flat: $(\rho+1)/2 \sim$ Beta$(1,1)$ & 120 vs 30 & 1.76 [0.87, 2.43] & 0.00 / 0.48 \\
Wickepin & Primary (saturated, N(0,1)) & 120 vs 30 & 1.75 [0.83, 2.43] & -0.04 / 0.28 \\
Wickepin & Wider: $\beta \sim \mathcal{N}(0, 2.5)$ & 120 vs 30 & 2.04 [1.21, 2.83] & -0.16 / 0.23 \\
Wickepin & Additive depth + variety & 120 vs 80 & 0.94 [0.16, 1.68] & 0.08 / 0.36 \\
Wickepin & Flat: $(\rho+1)/2 \sim$ Beta$(1,1)$ & 120 vs 80 & 0.95 [0.12, 1.67] & 0.00 / 0.48 \\
Wickepin & Primary (saturated, N(0,1)) & 120 vs 80 & 0.95 [0.16, 1.69] & -0.04 / 0.28 \\
Wickepin & Wider: $\beta \sim \mathcal{N}(0, 2.5)$ & 120 vs 80 & 1.03 [0.24, 1.82] & -0.16 / 0.23 \\
\bottomrule
\end{tabular}
\end{table}

\begin{table}[htbp]
\centering
\caption{Hazard ratios for halting development, from the sequential model with a complementary log-log link fitted on the principal-stage scale. Under this link the model is a grouped-time proportional hazards model whose event is the plant halting at its current stage, and $\exp(-\beta)$ is the proportional hazard ratio for that event. A ratio below one means the deeper treatment is less likely to stall at any given transition, that is, developmentally more advanced.}
\label{tab:hazard}
\small
\begin{tabular}{llrr}
\toprule
Site & Depth contrast (mm) & Hazard ratio [95\% CI] & P(HR $<1$) \\
\midrule
Merredin & 80 vs 30 & 0.22 [0.06, 0.86] & 0.981 \\
Merredin & 120 vs 30 & 0.23 [0.06, 0.86] & 0.982 \\
Merredin & 120 vs 80 & 1.19 [0.31, 3.22] & 0.502 \\
Wickepin & 80 vs 30 & 0.59 [0.35, 0.99] & 0.976 \\
Wickepin & 120 vs 30 & 0.32 [0.19, 0.54] & 0.999 \\
Wickepin & 120 vs 80 & 0.55 [0.32, 0.91] & 0.985 \\
\bottomrule
\end{tabular}
\end{table}

\begin{table}[htbp]
\centering
\caption{Between-site comparison of the variety-averaged sowing depth contrasts (Wickepin minus Merredin) on the latent logit scale. The two trials were fitted separately, so the posteriors are independent and their difference is obtained by pairing draws. $P(\text{same sign})$ is the posterior probability that the contrast has the same direction at both sites. This is a comparison of two separately fitted contrasts, not a joint multi-environment analysis.}
\label{tab:cross-site}
\small
\begin{tabular}{lrr}
\toprule
Depth contrast (mm) & Difference [95\% CI] & P(same sign) \\
\midrule
80 vs 30 & -0.61 [-2.54, 1.42] & 0.902 \\
120 vs 30 & 0.30 [-1.59, 2.33] & 0.936 \\
120 vs 80 & 0.91 [-1.03, 2.82] & 0.519 \\
\bottomrule
\end{tabular}
\end{table}

\begin{table}[htbp]
\centering
\caption{Variety--herbicide effects under the model of record, on the latent logit scale: each combination averaged over sowing depth and expressed as a deviation from the average combination, with 95\% credible intervals. A positive value means the combination is developmentally more advanced than the average.}
\label{tab:variety}
\small
\begin{tabular}{lrr}
\toprule
Variety--herbicide & Merredin & Wickepin \\
\midrule
V01 & -0.11 [-0.62, 0.39] & 0.17 [-0.25, 0.58] \\
V02 & -0.09 [-0.92, 0.75] & 0.06 [-0.50, 0.61] \\
V03 & -0.71 [-1.57, 0.14] & -0.21 [-0.75, 0.35] \\
V04 & 0.26 [-0.56, 1.06] & 0.48 [-0.04, 1.02] \\
V05 & -0.13 [-0.95, 0.70] & 0.10 [-0.45, 0.63] \\
V06 & 0.54 [-0.25, 1.33] & -0.86 [-1.42, -0.29] \\
V07 & 0.87 [0.08, 1.69] & 0.25 [-0.30, 0.79] \\
V08 & -0.51 [-1.35, 0.32] & 0.24 [-0.31, 0.79] \\
V09 & 0.29 [-0.52, 1.09] & -0.25 [-0.79, 0.29] \\
V10 & -0.39 [-1.21, 0.45] & 0.03 [-0.51, 0.58] \\
\bottomrule
\end{tabular}
\end{table}

\begin{table}[htbp]
\centering
\caption{Transition-specific parameters of the sequential model and the implied conditional probability of advancing past each stage, evaluated at $\eta = 0$. The design matrix is centred (Section~\ref{sec:fixed}), so this is the average treatment rather than a reference level, with all random effects at zero. Small advancement probabilities mark low conditional advancement at the census.}
\label{tab:thresholds}
\small
\begin{tabular}{llrr}
\toprule
Site & Transition & $\alpha_k$ [95\% CI] & P(advance) [95\% CI] \\
\midrule
Merredin & Z00$\to$Z01 & 6.54 [5.08, 8.18] & 0.998 [0.994, 1.000] \\
Merredin & Z01$\to$Z11 & 9.12 [6.94, 11.96] & 1.000 [0.999, 1.000] \\
Merredin & Z11$\to$Z12 & 0.61 [-0.65, 2.03] & 0.635 [0.343, 0.884] \\
Merredin & Z12$\to$Z13 & -1.69 [-2.97, -0.29] & 0.174 [0.049, 0.429] \\
Merredin & Z13$\to$Z21 & -3.94 [-5.30, -2.48] & 0.024 [0.005, 0.077] \\
Merredin & Z21$\to$Z22 & -6.07 [-8.82, -3.92] & 0.004 [0.000, 0.019] \\
Wickepin & Z11$\to$Z12 & 3.95 [3.26, 4.64] & 0.980 [0.963, 0.990] \\
Wickepin & Z12$\to$Z13 & 1.38 [0.76, 1.95] & 0.795 [0.682, 0.876] \\
Wickepin & Z13$\to$Z14 & -0.96 [-1.58, -0.37] & 0.280 [0.170, 0.408] \\
Wickepin & Z14$\to$Z15 & -1.31 [-1.98, -0.69] & 0.217 [0.121, 0.334] \\
Wickepin & Z15$\to$Z21 & 4.39 [2.67, 6.96] & 0.981 [0.935, 0.999] \\
Wickepin & Z21$\to$Z23 & -5.02 [-6.77, -3.67] & 0.009 [0.001, 0.025] \\
Wickepin & Z23$\to$Z24 & -1.03 [-3.89, 1.89] & 0.319 [0.020, 0.868] \\
\bottomrule
\end{tabular}
\end{table}

\begin{table}[htbp]
\centering
\caption{Posterior predictive probability that a plant in a new plot is at each Zadoks stage four weeks after sowing, averaged over the ten variety--herbicide combinations, with 95\% credible intervals. Predictions integrate over the replicate, main-plot, spatial and plot-level random effects.}
\label{tab:pred-probs}
\small
\begin{tabular}{lllll}
\toprule
Site & Stage & Sowing depth 30 mm & 80 mm & 120 mm \\
\midrule
Merredin & Z00 & 0.032 [0.006, 0.092] & 0.013 [0.001, 0.055] & 0.012 [0.001, 0.054] \\
Merredin & Z01 & 0.003 [0.000, 0.013] & 0.001 [0.000, 0.007] & 0.001 [0.000, 0.007] \\
Merredin & Z11 & 0.521 [0.301, 0.714] & 0.345 [0.180, 0.530] & 0.342 [0.164, 0.536] \\
Merredin & Z12 & 0.289 [0.183, 0.372] & 0.333 [0.238, 0.408] & 0.336 [0.238, 0.414] \\
Merredin & Z13 & 0.117 [0.027, 0.228] & 0.211 [0.109, 0.307] & 0.212 [0.105, 0.314] \\
Merredin & Z21 & 0.030 [0.001, 0.106] & 0.074 [0.021, 0.166] & 0.074 [0.018, 0.181] \\
Merredin & Z22 & 0.008 [0.000, 0.058] & 0.023 [0.001, 0.099] & 0.022 [0.001, 0.107] \\
Wickepin & Z11 & 0.074 [0.036, 0.141] & 0.038 [0.016, 0.086] & 0.016 [0.007, 0.041] \\
Wickepin & Z12 & 0.353 [0.246, 0.439] & 0.239 [0.156, 0.342] & 0.131 [0.078, 0.233] \\
Wickepin & Z13 & 0.430 [0.340, 0.489] & 0.458 [0.366, 0.513] & 0.408 [0.318, 0.475] \\
Wickepin & Z14 & 0.099 [0.058, 0.152] & 0.155 [0.103, 0.202] & 0.205 [0.149, 0.248] \\
Wickepin & Z15 & 0.000 [0.000, 0.002] & 0.001 [0.000, 0.003] & 0.001 [0.000, 0.004] \\
Wickepin & Z21 & 0.042 [0.014, 0.110] & 0.103 [0.047, 0.201] & 0.224 [0.123, 0.341] \\
Wickepin & Z23 & 0.000 [0.000, 0.002] & 0.002 [0.000, 0.007] & 0.004 [0.000, 0.018] \\
Wickepin & Z24 & 0.001 [0.000, 0.006] & 0.004 [0.000, 0.018] & 0.012 [0.001, 0.048] \\
\bottomrule
\end{tabular}
\end{table}

\begin{table}[htbp]
\centering
\caption{Convergence diagnostics for every model fitted. $\hat{R}$ is the rank-normalised split-$\hat{R}$ and the effective sample sizes are the rank-normalised bulk and tail ESS of \citet{Vehtari2021rhat}, taken over all thresholds, fixed effects, variance components and autocorrelation parameters.}
\label{tab:diagnostics}
\resizebox{\textwidth}{!}{%
\begin{tabular}{lllllrrrr}
\toprule
Site & Stage & Model & Family & Fixed & max $\hat{R}$ & min bulk ESS & min tail ESS & Divergences \\
\midrule
Merredin & 1\_random & R0: rep + main-plot & sratio & interaction & 1.001 & 5013 & 6173 & 1 \\
Merredin & 1\_random & R1: R0 + plot nugget & sratio & interaction & 1.0022 & 4395 & 3570 & 2 \\
Merredin & 1\_random & R2: R0 + AR1$\times$AR1 & sratio & interaction & 1.0009 & 5119 & 4906 & 0 \\
Merredin & 1\_random & R3: R0 + AR1$\times$AR1 + nugget & sratio & interaction & 1.0032 & 1036 & 1980 & 0 \\
Merredin & 1\_random & R4: AR1$\times$AR1 + nugget, no blocking & sratio & interaction & 1.0024 & 1571 & 2827 & 0 \\
Merredin & 2\_family & Cumulative & cumulative & interaction & 1.0027 & 1095 & 1696 & 0 \\
Merredin & 2\_family & Adjacent-category & acat & interaction & 1.0011 & 1786 & 3295 & 0 \\
Merredin & 3\_fixed & Depth only & sratio & depth & 1.0034 & 937 & 2248 & 2 \\
Merredin & 3\_fixed & Depth + variety & sratio & additive & 1.002 & 932 & 2008 & 0 \\
Merredin & 4\_sensitivity & S\_widebeta & sratio & interaction & 1.0021 & 1130 & 1892 & 0 \\
Merredin & 4\_sensitivity & S\_flatrho & sratio & interaction & 1.0011 & 1276 & 1393 & 1 \\
Wickepin & 1\_random & R0: rep + main-plot & sratio & interaction & 1.001 & 5598 & 4886 & 6 \\
Wickepin & 1\_random & R1: R0 + plot nugget & sratio & interaction & 1.0008 & 3435 & 3543 & 2 \\
Wickepin & 1\_random & R2: R0 + AR1$\times$AR1 & sratio & interaction & 1.0009 & 4011 & 4698 & 0 \\
Wickepin & 1\_random & R3: R0 + AR1$\times$AR1 + nugget & sratio & interaction & 1.0054 & 1237 & 2330 & 0 \\
Wickepin & 1\_random & R4: AR1$\times$AR1 + nugget, no blocking & sratio & interaction & 1.0014 & 1103 & 1980 & 0 \\
Wickepin & 2\_family & Cumulative & cumulative & interaction & 1.0016 & 1616 & 2535 & 2 \\
Wickepin & 2\_family & Adjacent-category & acat & interaction & 1.003 & 1525 & 3303 & 3 \\
Wickepin & 3\_fixed & Depth only & sratio & depth & 1.0096 & 727 & 1040 & 0 \\
Wickepin & 3\_fixed & Depth + variety & sratio & additive & 1.0045 & 1057 & 1552 & 0 \\
Wickepin & 4\_sensitivity & S\_widebeta & sratio & interaction & 1.0009 & 1294 & 2700 & 0 \\
Wickepin & 4\_sensitivity & S\_flatrho & sratio & interaction & 1.0012 & 1668 & 1530 & 0 \\
\bottomrule
\end{tabular}%
}
\end{table}

\begin{figure}[htbp]
\centering
\includegraphics[width=0.92\linewidth]{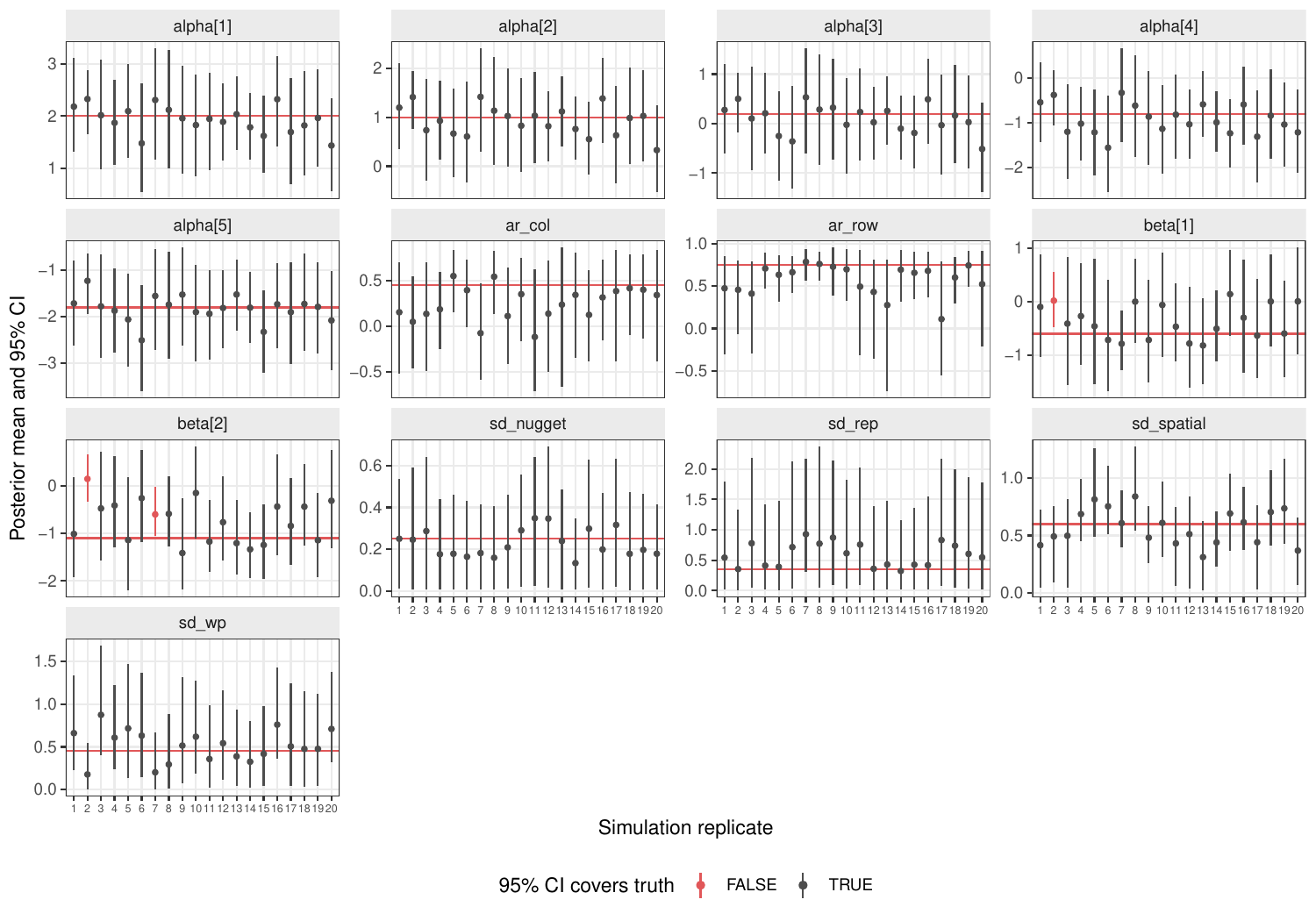}
\caption{Parameter recovery. Twenty data sets were simulated from the full model on the real 5\,$\times$\,18 field layout with ten plants per plot and refitted. Points and intervals are posterior means and 95\% credible intervals; horizontal red lines are the values used to simulate. Intervals that fail to cover the truth are highlighted.}
\label{fig:recovery}
\end{figure}

\begin{figure}[htbp]
\centering
\includegraphics[width=0.92\linewidth]{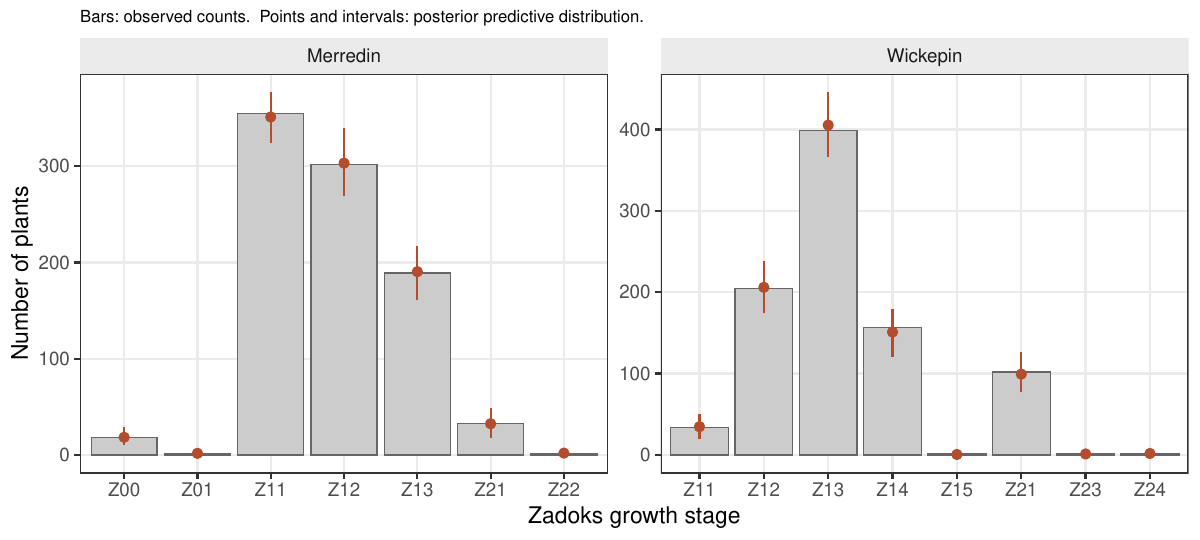}
\caption{Posterior predictive check. Bars are the observed number of plants at each stage; points and intervals are the posterior predictive distribution under the model of record.}
\label{fig:ppc}
\end{figure}

\begin{figure}[htbp]
\centering
\includegraphics[width=0.92\linewidth]{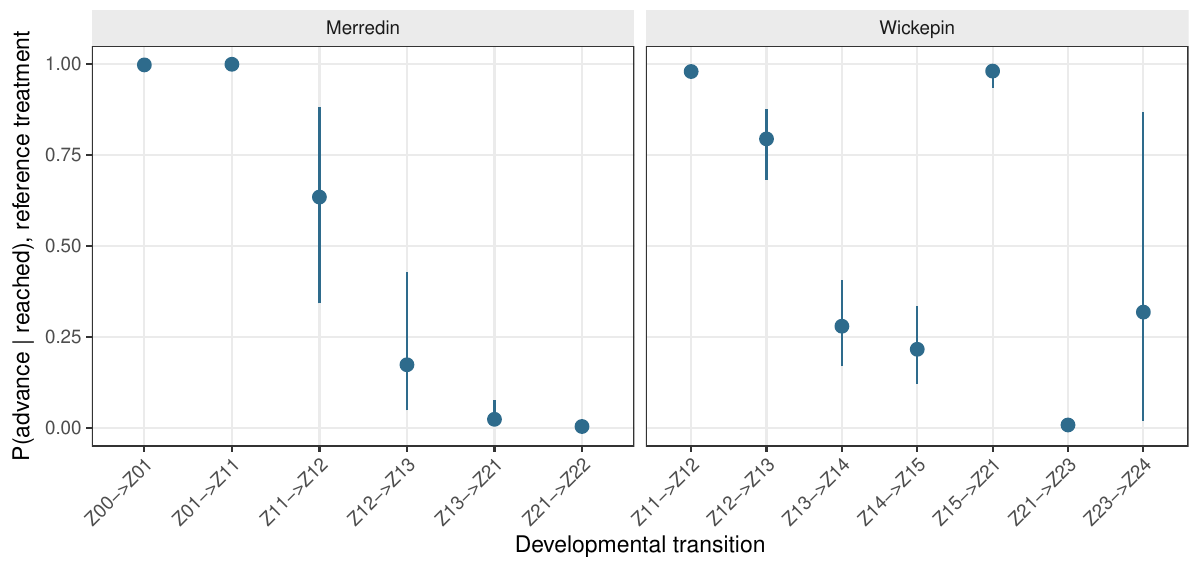}
\caption{Conditional probability of advancing past each stage under the sequential model, evaluated at $\eta = 0$. Low values mark developmental bottlenecks. The monotone decline is expected from a single census of a progressing population, and the estimates are imprecise where a category was recorded only once or twice.}
\label{fig:thresholds}
\end{figure}

\begin{figure}[htbp]
\centering
\includegraphics[width=0.92\linewidth]{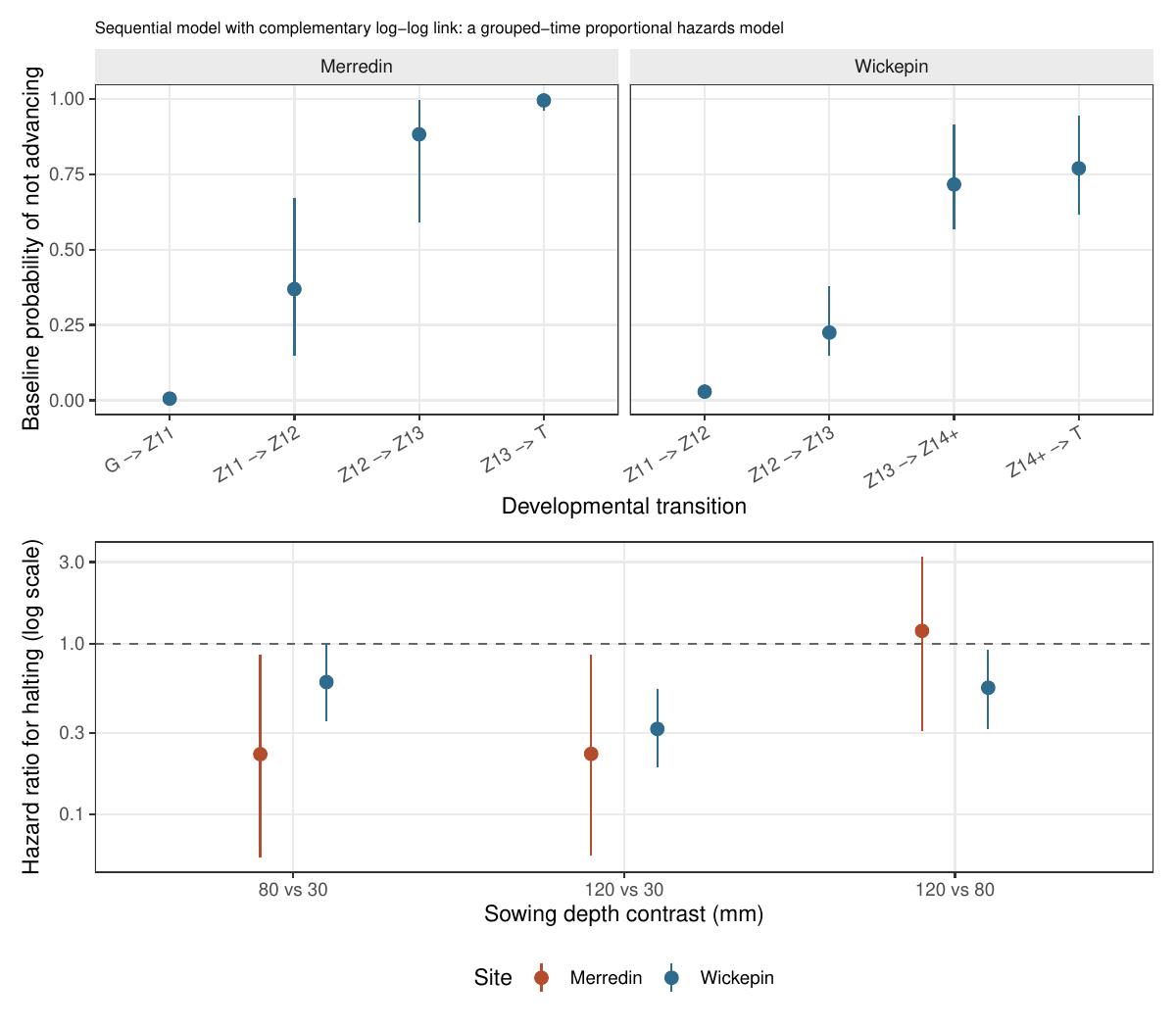}
\caption{The sequential model with a complementary log-log link, a grouped-time proportional hazards model whose event is the plant halting at its current stage, fitted on the principal-stage scale. Upper: the baseline probability of not advancing past each stage, at the average treatment. Lower: hazard ratios for halting under deeper sowing; values below one indicate development that is less likely to stall, and so more advanced.}
\label{fig:hazard}
\end{figure}

\begin{figure}[htbp]
\centering
\includegraphics[width=0.92\linewidth]{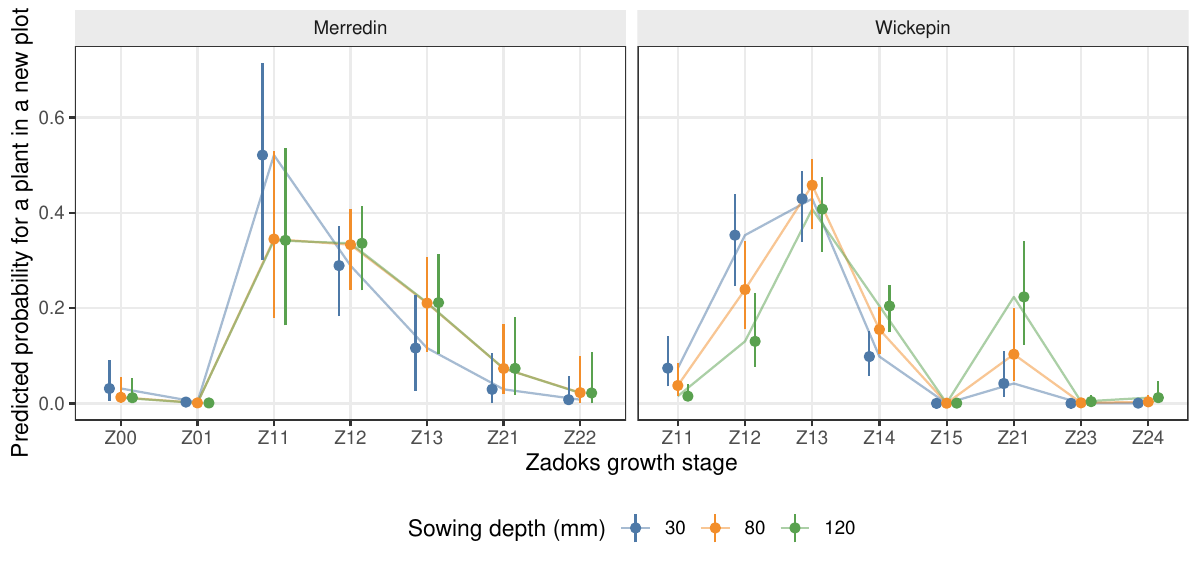}
\caption{Posterior predictive probability that a plant in a new plot is at each Zadoks stage four weeks after sowing, by sowing depth, averaged over the ten variety--herbicide combinations. Intervals are 95\% credible intervals.}
\label{fig:predprobs}
\end{figure}

\section{Zadoks growth scale reference}
\label{app:zadoks}

The Zadoks decimal growth scale \citep{Zadoks1974} describes cereal development with a two-digit code in which the first digit denotes the principal stage and the second the secondary stage within it. The stages relevant to this study are:

\paragraph{Stage 0: germination}
\begin{itemize}[noitemsep]
\item Z00: Dry seed
\item Z01: Start of water absorption
\end{itemize}

\paragraph{Stage 1: seedling growth}
Leaves are counted on the main stem only; a leaf is ``emerged'' when the ligule is visible.
\begin{itemize}[noitemsep]
\item Z11: First leaf emerged
\item Z12: Two leaves emerged
\item Z13: Three leaves emerged
\item Z14: Four leaves emerged
\item Z15: Five leaves emerged
\end{itemize}

\paragraph{Stage 2: tillering}
Visible tillers are counted, a tiller being a side shoot with a leaf blade emerging between a leaf sheath and the main stem.
\begin{itemize}[noitemsep]
\item Z21: Main stem and one tiller
\item Z22: Main stem and two tillers
\item Z23: Main stem and three tillers
\item Z24: Main stem and four tillers
\end{itemize}

\noindent The scale is not monotone in calendar time. A crop at Z13 will initiate tillering (Z21) before completing its fourth leaf (Z14), which is the non-standard Z13~$\to$~Z21 transition modelled here.

\clearpage

\end{document}